\documentclass[reprint, superscriptaddress,nofootinbib]{revtex4-1}

\usepackage{graphicx}
\usepackage{color}

\usepackage{amsmath}
\usepackage{amsfonts}
\usepackage{amssymb}

\newcommand{\relampSTD}{20\%}
\newcommand{\relphsSTD}{\ensuremath{15^\circ}}
\newcommand{\relampSTDem}{15\%}

\newcommand{\sharedabsampSTD}{25\%}
\newcommand{\sharedabsphsSTD}{\ensuremath{27^\circ}}

\newcommand{\relnumTEN}{10--20}
\newcommand{\relnumONE}{1000--2000}

\newcommand{\absnumONEem}{400--600}

\newcommand{\limnum}{2--15\%}

\begin{document}

\title{
Calibrating gravitational-wave detectors with GW170817
}

\author{
Reed Essick and Daniel E. Holz \\
\small{Kavli Institute for Cosmological Physics} \\
\small{The University of Chicago, 5640 South Ellis Avenue, Chicago, Illinois, 60637, USA}
}

\begin{abstract}
The waveform of a compact binary coalescence is predicted by general relativity.
It is therefore possible to directly constrain the response of a gravitational-wave (GW) detector by analyzing a signal's observed amplitude and phase evolution as a function of frequency.
GW signals alone constrain the relative amplitude and phase between different frequencies within the same detector and between different detectors.
Furthermore, if the source's distance and inclination can be determined independently, for example from an electromagnetic counterpart, one can calibrate the absolute amplitude response of the detector network.
We analyze GW170817's ability to calibrate the LIGO/Virgo detectors, finding a relative amplitude calibration precision of approximately $\pm\relampSTD$ and relative phase precision of $\pm\relphsSTD$ (1-$\sigma$ uncertainty) between the LIGO Hanford and Livingston detectors.
Incorporating additional information about the distance and inclination of the source from electromagnetic observations, the relative amplitude of the LIGO detectors can be tightened to $\sim\pm\relampSTDem$.
Including electromagnetic observations also constrains the absolute amplitude precision to similar levels.
We investigate the ability of future events to improve astronomical calibration.
By simulating the cumulative uncertainties from an ensemble of detections, we find that with several hundred events with  electromagnetic counterparts, or several thousand events without counterparts, we reach percent-level astronomical calibration.
This corresponds to $\sim$5--10 years of operation at advanced LIGO and Virgo design sensitivity.
It is to be emphasized that direct {\em in-situ}\/ measurements of detector calibration provide significantly higher precision than astronomical sources, and already constrain the calibration to a few percent in amplitude and a few degrees in phase.
In this sense, our astronomical calibrators only corroborate existing calibration measurements.
Nonetheless, it is remarkable that we are able to use an astronomical GW source to characterize properties of a terrestrial GW instrument, and astrophysical calibration may become an important corroboration of existing calibration methods, providing a completely independent constraint of potential systematics.
\end{abstract}

\maketitle

\section{Introduction}
\label{section:introduction}

Gravitational-wave (GW) astronomy has become an important new discipline within modern astrophysics.
In particular, the LIGO/Virgo detection of the coalescence of two neutron stars (GW170817; \cite{GW170817}) constrained the supra-nuclear equation of state within neutron star cores~\cite{SourceProperties,EOS,Landry2018} and multi-messenger observations of the associated electromagnetic (EM) counterparts in the host galaxy NGC~4993 (GRB170817a~\cite{GRB} and AT 2017gf0~\cite{multimessenger}) informed GRB and kilonova phenomenology (see, e.g.,~\cite{kilonova,Coughlin2018}) and provided a direct standard-siren measurement of the Hubble constant ($H_0$; \cite{H0}).
Although the full breadth of the science enabled by the advanced LIGO~\cite{LIGO} and Virgo~\cite{Virgo} detectors is too long to enumerate here (see~\cite{Catalog, Populations} for summaries), we note that these discoveries were possible only with both accurate and precise interferometric calibration~\cite{PhysRevD.95.062003, PhysRevD.96.102001, 0264-9381-35-20-205004}.
Because most, if not all, inferences made with GW data compare the observed signal to theoretical predictions, it is vital that the recorded data is correctly calibrated.
Errors in calibration of the detectors will result in errors in derived parameters, including distance (due to amplitude errors) and sky position (due to relative timing, phase, and amplitude errors).
Indeed, accurate high-precision calibration is an essential component of GW astronomy, and in what follows we focus on an independent method for characterizing the precision of GW instruments.

Calibration uncertainty in the current network of GW detectors was limited to a few percent in amplitude and a few degrees in phase during the first two observing runs~\cite{PhysRevD.96.102001,PhysRevD.95.062003,0264-9381-35-20-205004}.
While the uncertainty in detectors' responses has roughly improved with their sensitivity to date, existing calibration techniques are quickly approaching their fundamental limits.
These techniques are based upon either phyiscally displacing the detectors' test masses, or causing an equivalent displacements, by independent ``first-principles'' systems with verifiable accuracy and high precision.
Many such fundamental references have been used, including the primary laser's wavelength, as was done for initial LIGO \cite{ABADIE2010223} and advanced VIRGO \cite{0264-9381-35-20-205004}, auxiliary laser systems currently used by advanced LIGO, such as photon calibrators \cite{Karki2016}, and corroborating techniques using the detectors' frequency stabilization system \cite{Goetz_2010, PhysRevD.95.062003}, or local gravity gradient generators~\cite{2018arXiv180606572E}.
Each of these systems are used either directly or indirectly to induce strain within the detector and thus measure the uncertainty in the detector's response.
Each is fundamentally limited by the uncertainty in its reference.

There exists another, completely independent, way to induce strain on a GW detector: astrophysical sources.
If we assume that the theory of gravity is known to sufficient precision (i.e., general relativity accurately describes our sources and GW propagation), we can use observed astrophysical sources to infer the calibration of each detector within the global network.
The basic idea behind astrophysical calibration is that the amplitude and phase evolution of a detected source is constrained by general relativity.
By carefully analyzing the detected waveform, it is possible to directly measure how the detector responds to impinging gravitational waves.
There are a range of potential constraints, including i) the relative phase and amplitude, as a function of frequency, for a single detector, ii) the relative phase and amplitude between multiple detectors, and iii) the absolute amplitude and phase calibration of a detector network.
We explore all of these in turn.
We emphasize that the remarkable accuracy and precision of existing {\em in-situ}\/ calibration techniques are essential to LIGO/Virgo science (see, for example, \cite{SourceProperties,EOS,H0,NLTides}).
Astronomical calibration will primarily provide an independent corroboration of these techniques.

This idea was first proposed in a prescient paper by Pitkin, Messenger, and Wright~\cite{Pitkin2016}.
They focused on the possibility of constraining the overall amplitude calibration, and argued that a BNS within 100\,Mpc, with an EM counterpart that somewhat optimistically determines the distance to the source exactly and places a strong prior on the inclination, would constrain the amplitude response to 10\%. 
We extend their work, focusing on both absolute and relative responses of GW detectors.
In addition, we generalize their approach, and consider both amplitude and phase errors, and incorporate possible frequency dependence.
Furthermore, we apply our approach to real data, deriving constraints on systematic errors from the observation of GW170817.
For GW170817, we apply conservative priors based on EM observations and find similarly looser constraints on detector calibration.
Finally, similar to~\cite{Pitkin2016}, we also perform a projection of future constraints.

We describe our formalism in \S\ref{section:formalism}, and then derive calibration constraints from the observation of GW170817 in~\S\ref{section:results}.
We also describe the key features of astronomical calibration and how these depend on the nature of the sources (e.g., GW170817 was near a null in Virgo's antenna response; more favorable source locations may result in better polarization measurements, and therefore better inclination constraints).
We estimate how astronomical calibration scales with an ensemble of detections in~\S\ref{section:scaling} and conclude in~\S\ref{section:conclusions}.

\section{Formalism}
\label{section:formalism}

We assume additive stationary Gaussian noise ($n$) in each detector and model the data recorded ($d$) as a function of frequency as
\begin{equation}\label{model}
    d = n + \left(1+\delta A\right)e^{i\delta \psi} \left(F_{+} h_+ + F_{\times} h_{\times}\right)
\end{equation}
where $\delta A$ is the amplitude calibration error, $\delta \psi$ the phase error, $F_{+,\times}$ are the antenna response functions,\footnote{For current ground-based detectors, the antenna response functions are approximately frequency-independent, and we make this assumption here. However, future detectors, with significantly longer arms~\cite{CosmicExplorer,EinsteinTelescope}, may violate this assumption~\cite{FreqDepAntennaResponse}.} and $h_{+,\times}$ the astrophysical strain incident on the detectors.
The strain incident on the detector can be written as
\begin{align}
    h_+ = \frac{1+\cos^2\theta_\mathrm{jn}}{2 D_L} e^{i\phi_o} h           \label{plus_strain} \\
    h_\times = \frac{\cos\theta_\mathrm{jn}}{D_L} e^{i(\phi_o + \pi/2)} h \label{cross_strain}
\end{align}
where $D_L$ is the luminosity distance to the source, $\theta_\mathrm{jn}$ the angle between the source's total angular momentum and our line of sight, $\phi_o$ the orbital phase at coalescence, and $h$ the intrinsic waveform generated by the source.
If the recorded data's calibration is correct, then $\delta A=\delta \psi=0$.
Calibration errors are modeled as a function of frequency by spline interpolation, as implemented by \texttt{LALInference}~\cite{LALInference, lalsuite}.
Anchor points are logarithmically spaced in frequency separately for both amplitude and phase errors, and the interpolation is performed with the log-frequency as the abscissa.
The values from the posterior for the calibration errors in each detector at these anchor points are sampled simultaneously with the rest of the source's parameters.
This spline approach, albeit with significantly tighter priors on the calibration uncertainty derived from direct measurements of the detector response, has been used extensively in LIGO/Virgo analyses (e.g.,~\cite{SourceProperties,EOS,H0,NLTides}).

As we detail below, there are a number of challenges associated with using GW signals to calibrate GW detectors.
Measurement uncertainties between calibration errors at different frequencies are correlated within each detector and between detectors by the fact that all detectors observe the same signal.
In particular, $D_L$ and $\theta_\mathrm{jn}$ correlate with the absolute amplitude and $\phi_o$ correlates with the absolute phase.
Inference based on GW data alone can only constrain the relative amplitude and phase, both between different frequencies in each detector and between detectors.

Furthermore, due to the three-fold degeneracy between $\delta A$, $D_L$, and $\theta_\mathrm{jn}$, using EM constraints for either $D_L$ or $\theta_\mathrm{jn}$ alone does not significantly reduce the posterior uncertainty for $\delta A$; only when EM constraints for both $D_L$ and $\theta_\mathrm{jn}$ are imposed do we see a significant increase in $\delta A$'s precision.
Furthermore, EM information constrains the absolute calibration amplitude uncertainty, and therefore the amplitude uncertainty in each detector can be separately bounded.
Constraints on $D_L$ and $\theta_\mathrm{jn}$ do not significantly affect the uncertainty in $\delta \psi$, much like they do not affect inference of other intrinsic parameters~\cite{Pankow2017}.
This is because EM data does not observe the overall phase, whereas it does observe the overall amplitude.

\section{Calibration from GW170817}
\label{section:results}

\begin{figure*}
    \includegraphics[width=0.49\textwidth]{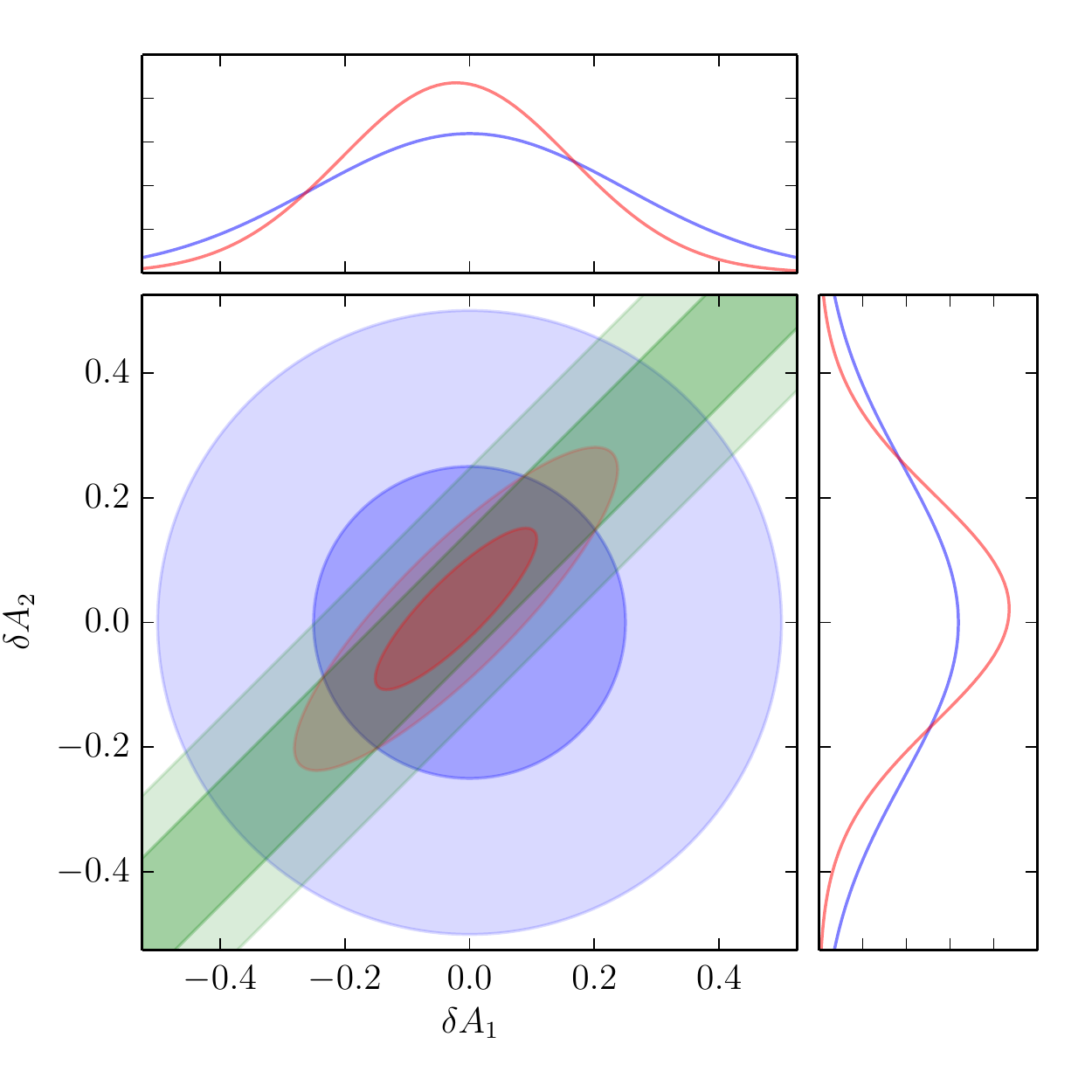}
    \includegraphics[width=0.49\textwidth]{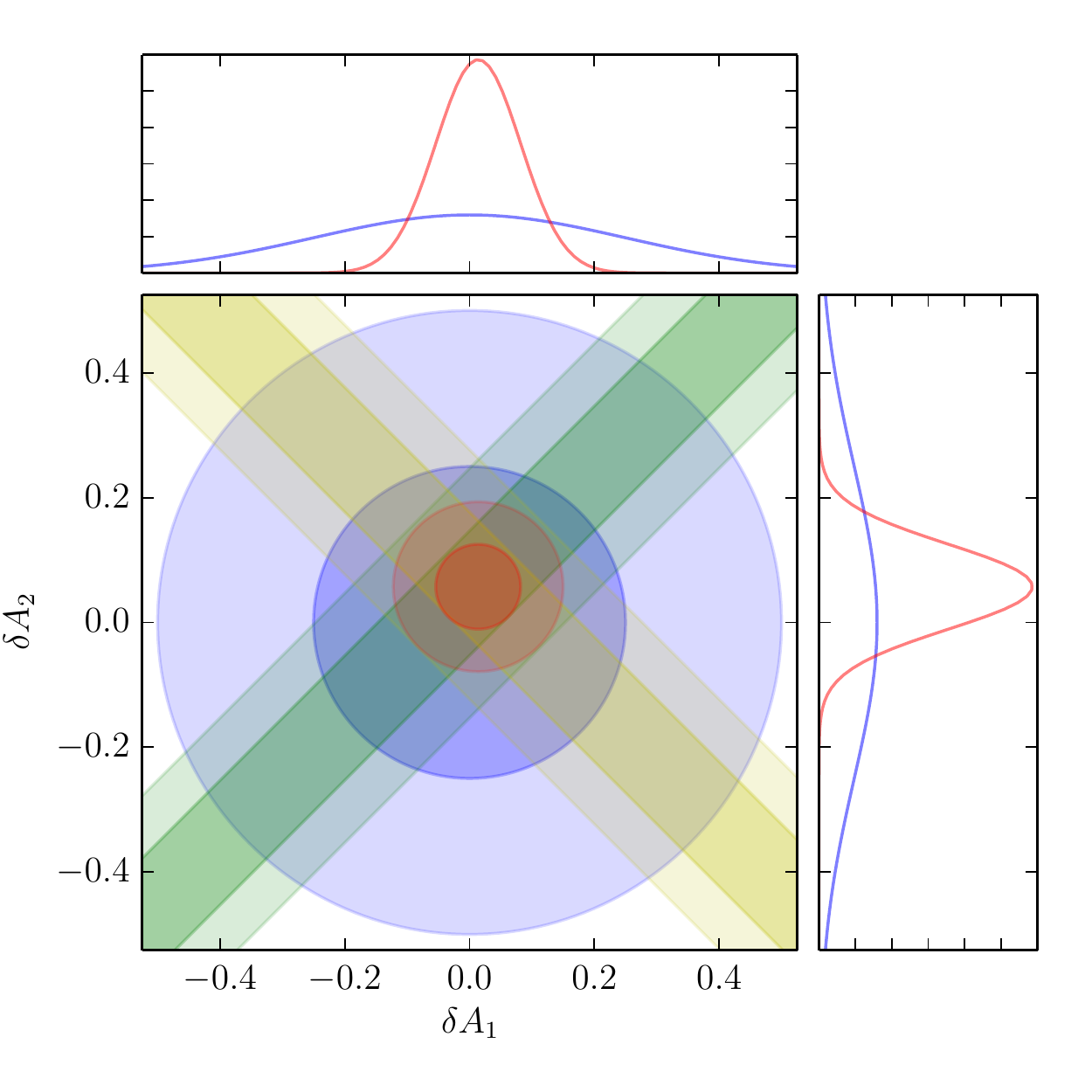}
    \caption{
        (\textit{left}) Toy model showing contraints obtained with only GW data with our prior ({blue}), likelihood ({green}), and posterior ({red}).
        The GW likelihood only constraints the relative amplitude (the difference) but not the overall amplitude (the sum).
        (\textit{right}) The same toy model showing how EM constraints ({yellow}) on $D_L$ and $\theta_\mathrm{jn}$ can further constrain the amplitude.
        Knoweldge of both $D_L$ and $\theta_\mathrm{jn}$ primarily constrains the overall sum.
        This behavior is born out in Figure~\ref{figure:amp corner}.
        See Appendix~\ref{section:toy model} for more details.
    }
    \label{figure:cartoon}
\end{figure*}

We re-analyze GW170817 with loose priors on the detectors' calibration.
Using \texttt{LALInference}~\cite{LALInference} (publicly available within \texttt{LALSuite}~\cite{lalsuite}), standard power-spectral-density estimation techniques, and a \texttt{TaylorF2} frequency-domain waveform that includes the effects of aligned spins and linear tides (see the review in~\cite{Buonanno2009}), we investigate the extent to which the astrophysical strain from a reasonably well-known source can be used to calibrate our detectors.

\begin{figure*}
    \begin{minipage}{0.49\textwidth}
        \begin{center}
            \includegraphics[width=1.0\columnwidth, clip=True, trim=0.65cm 1.10cm 0.28cm 0.00cm]{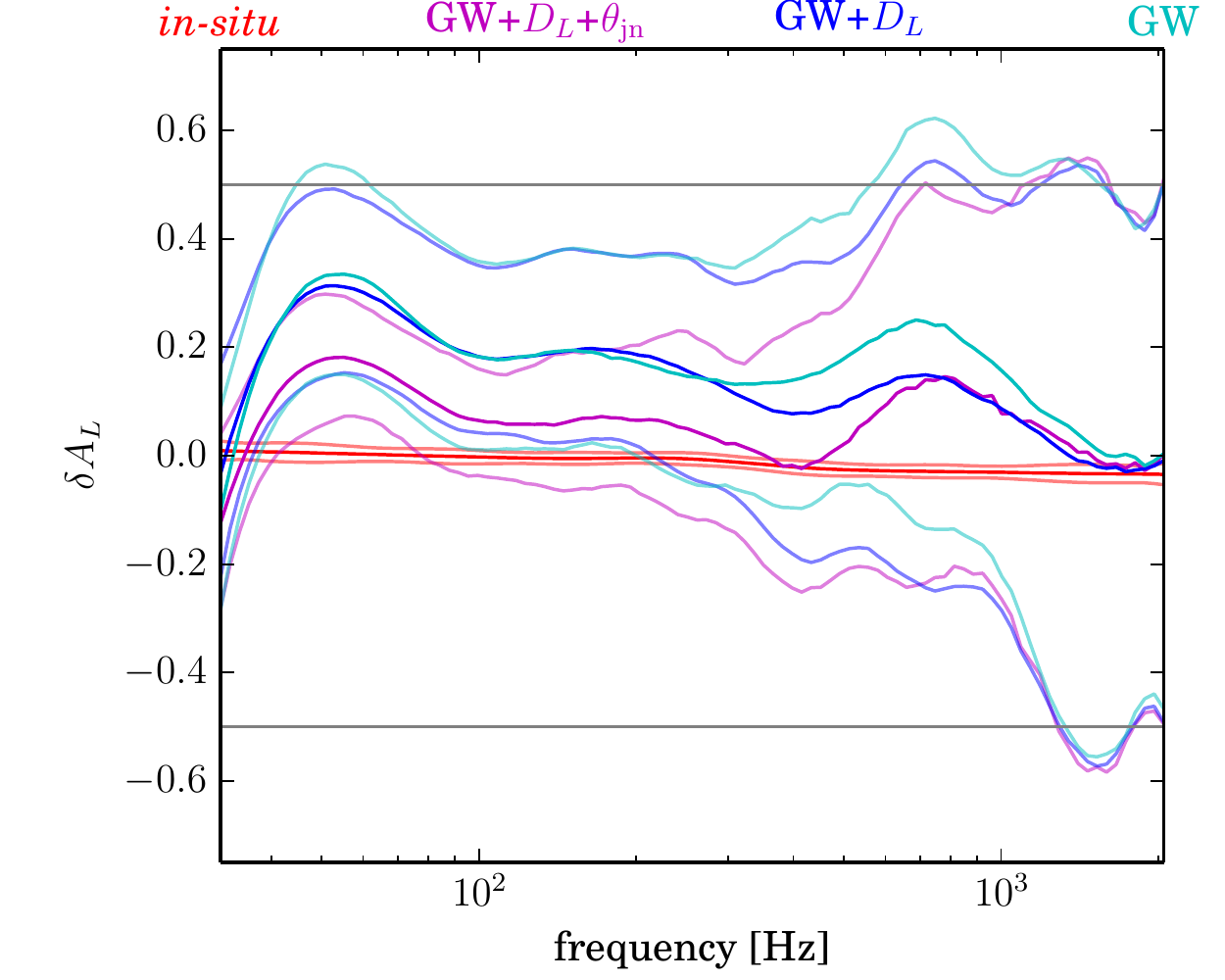} \\
            \includegraphics[width=1.0\columnwidth, clip=True, trim=0.65cm 0.05cm 0.28cm 0.50cm]{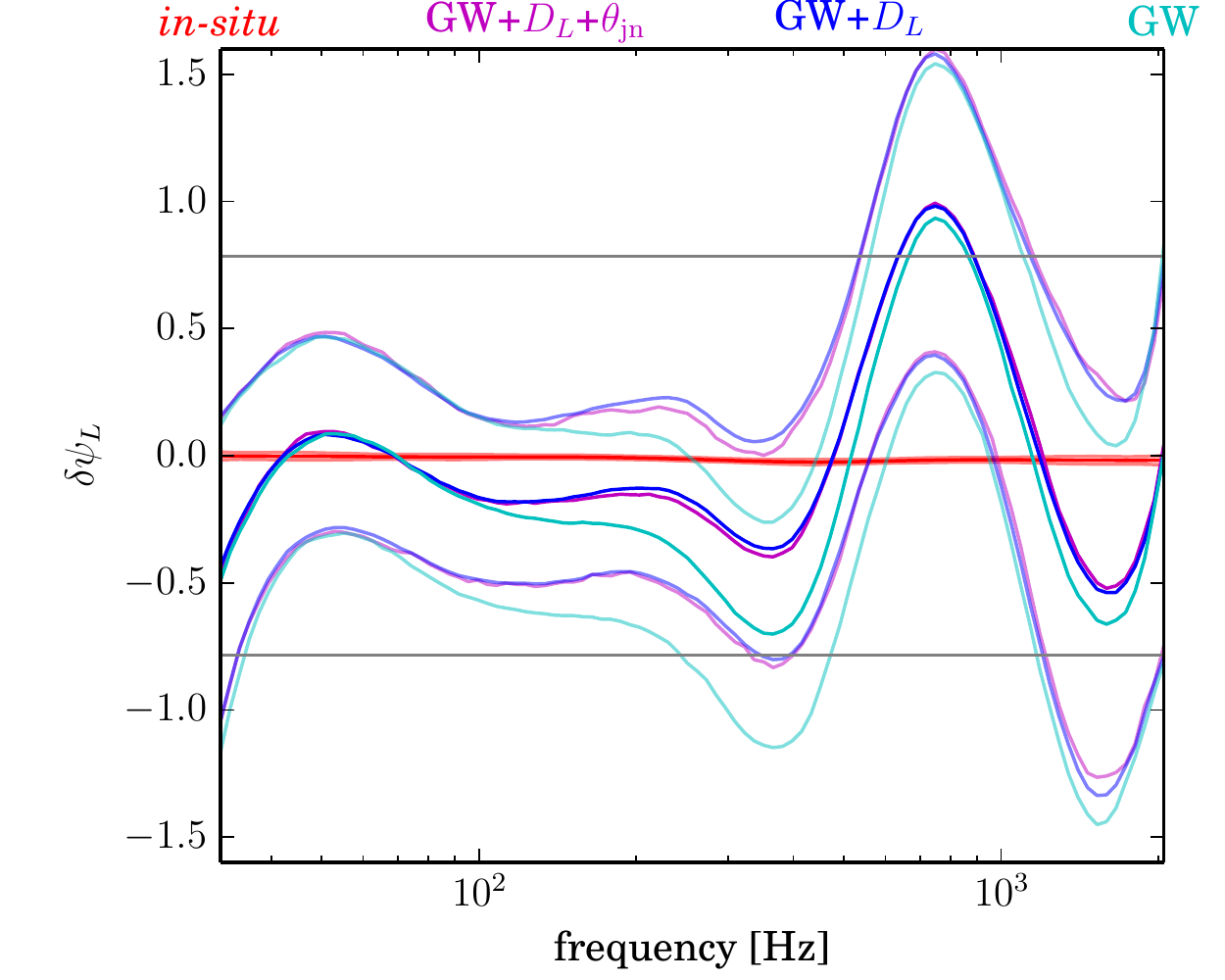} \\
        \end{center}
    \end{minipage}
    \begin{minipage}{0.49\textwidth}
        \begin{center}
            \includegraphics[width=1.0\columnwidth, clip=True, trim=0.65cm 1.10cm 0.28cm 0.00cm]{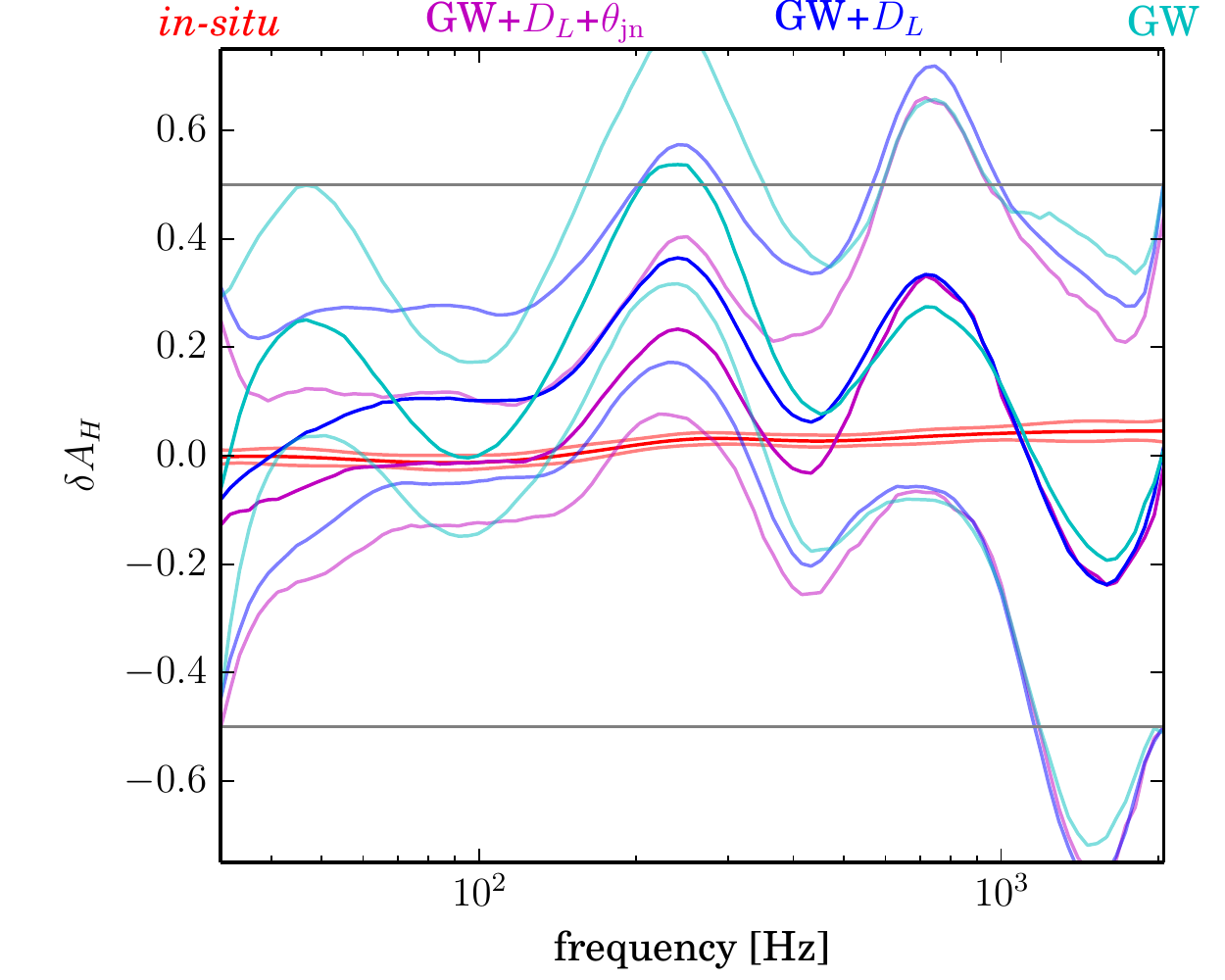} \\
            \includegraphics[width=1.0\columnwidth, clip=True, trim=0.65cm 0.05cm 0.28cm 0.50cm]{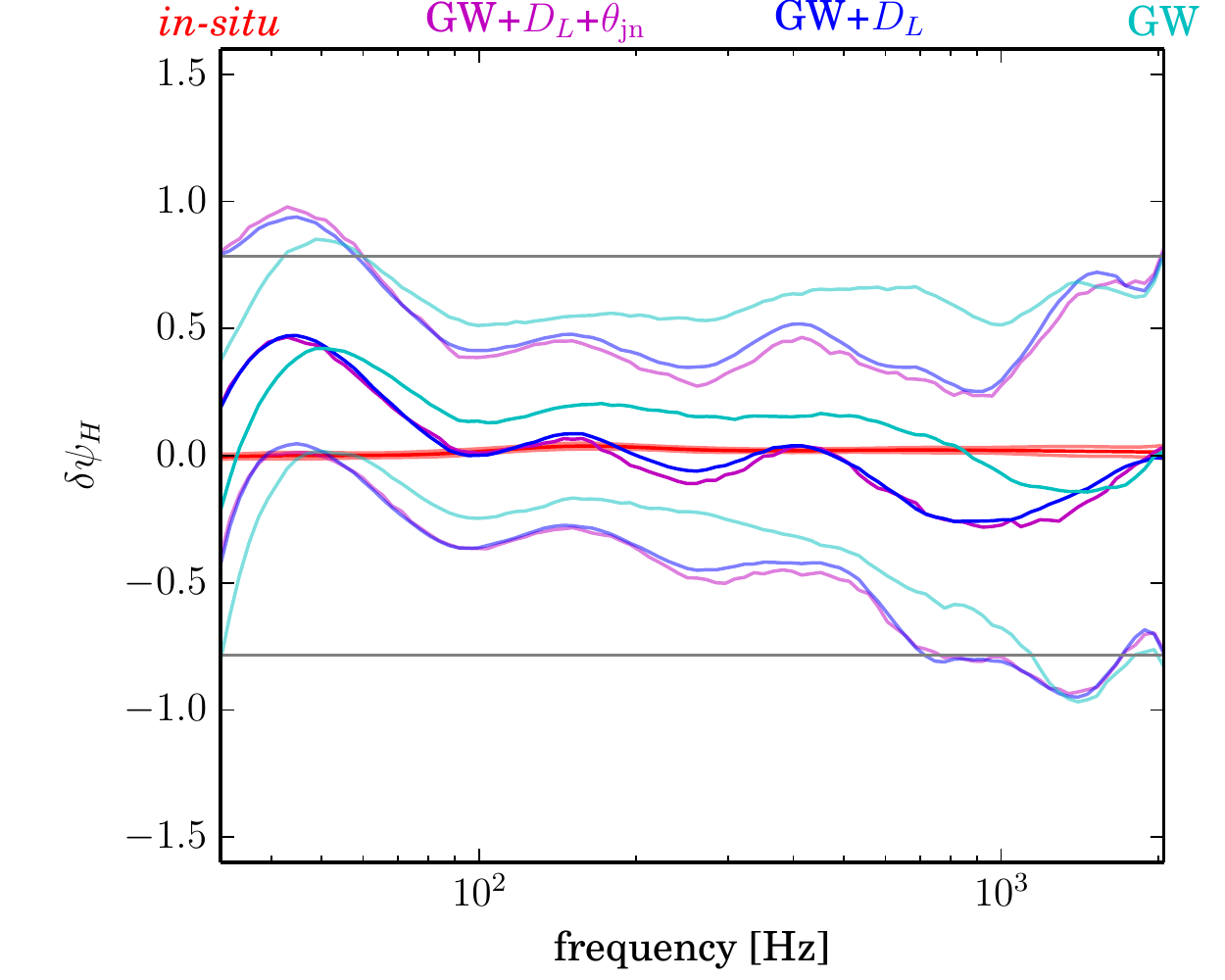} \\
        \end{center}
    \end{minipage}
    \caption{
        Posterior processes for both absolute amplitude and phase calibration errors for the LIGO Livingston (\emph{left}) and LIGO Hanford (\emph{right}) detectors.
        Dark colored lines correspond to the median \textit{a posteriori}; light colored lines correspond to the 1-$\sigma$ \textit{posterior} credible regions; grey lines correspond to 1-$\sigma$ \textit{prior} credible regions.
        We include results using the directly-measured calibration uncertainties (red: \textit{in-situ}), wide calibration prior uncertainties with only GW data ({light blue}: GW), wide calibration priors and EM constraints on $D_L$ ({dark blue}: GW+$D_L$), and wide calibration priors and EM constraints on both $D_L$ and $\theta_\mathrm{jn}$ ({purple}: GW+$D_L$+$\theta_\mathrm{jn}$).
    }
    \label{figure:processes}
\end{figure*}

Previous analyses used priors based on direct measurements of detector calibrations~\cite{PhysRevD.96.102001} with careful quantification of the uncertainty in both $\delta A$ and $\delta \psi$ as a function of frequency~\cite{SourceProperties,EOS,H0,NLTides}.
Like this work, these analyses model the calibration uncertainties as a function of frequency via spline interpolation between logarithmically spaced anchor points.
Priors for $\delta A$ and $\delta \psi$ at each anchor point are independent Gaussians with means and standard deviations taken from the measured uncertainties, typically corresponding to uncertainties of a few percent in amplitude and a few degrees in phase (see Figure~\ref{figure:processes}).
As we will show, these \textit{in-situ}\/ calibration measurements are significantly more precise than the constraints inferred from GW170817.

In addition to analyzing GW170817 with directly-measured calibration uncertainties, we impose independent zero-mean Gaussian priors with standard deviations significantly wider than would ever be encountered in science-quality data.
To wit, we assume loose \textit{a priori} 1-$\sigma$ calibration precision of $50\%$ in amplitude and $45^\circ$ in phase.
As we will see, GW170817 constrains the precision to be roughly a factor of two better \textit{a posteriori}, and does so with a non-trivial frequency dependence.

While the detection of the EM counterpart to GW170817, AT 2017gfo, determines the source's location to high precision, we find that the three-interferometer LIGO+Virgo network already provides sufficient precision in GW170817's right ascension ($\alpha$) and declination ($\delta$) such that additional refinements in ($\alpha$, $\delta$) do not significantly affect our results.
That is to say, the three-detector network limits the source's location to a small enough region that the antenna response functions do not vary significantly over the posterior's support.
Therefore, the impact of refined localization on the projected strain within each detector is relatively small.
This was observed empirically, as the posteriors for our calibration parameters were virtually unchanged when we fixed ($\alpha$, $\delta$) to the known values for NGC 4993~\cite{multimessenger}.
Indeed, previous studies have shown that precise localization does not necessarily improve inference of other parameters~\cite{Pankow2017}.

We find interesting constraints \textit{a posteriori}\/ only when the projected strain in a given detector is above the noise floor ($\sim$40--300 Hz, see Figure 1 of~\cite{GW170817}). The calibration at high or low frequencies remains uninformed by the data.
This is expected, since the additional freedom allowed by frequency-dependent calibration uncertainty gives the signal model less constraining power.
Our inference therefore acts like an unmodeled transient search, in which coherence between detectors and projected strains larger than the noise determine where the data is most informative.

The combination of GW and EM data for a given source allows for qualitatively different calibration constraints.
As mentioned in the previous section, the degeneracy between $\delta A$, $D_L$, and $\theta_\mathrm{jn}$, means that GW data alone cannot inform the absolute amplitude calibration.
Instead, knowledge of the signal's frequency and amplitude evolution informs the relative amplitude calibration, as shown in Figure~\ref{figure:cartoon} and Appendix~\ref{section:toy model}.\footnote{Similar arguments exist for $\delta \psi$, although EM data does not constrain $\phi_o$ and therefore does not affect our inference of the absolute phase.}
Conversely, EM constraints on both $D_L$ and $\theta_\mathrm{jn}$ primarily inform the absolute amplitude, typically limited by uncertainty in $H_0$ from $D_L$ and relativistic ejecta dynamics and energetics for $\theta_{jn}$, which acts orthogonally to the GW data.
Figure~\ref{figure:cartoon} shows how both GW and EM data combine to constrain $\delta A$.

\begin{figure}
    \includegraphics[width=1.0\columnwidth, clip=True, trim=0.65cm 1.10cm 0.28cm 0.00cm]{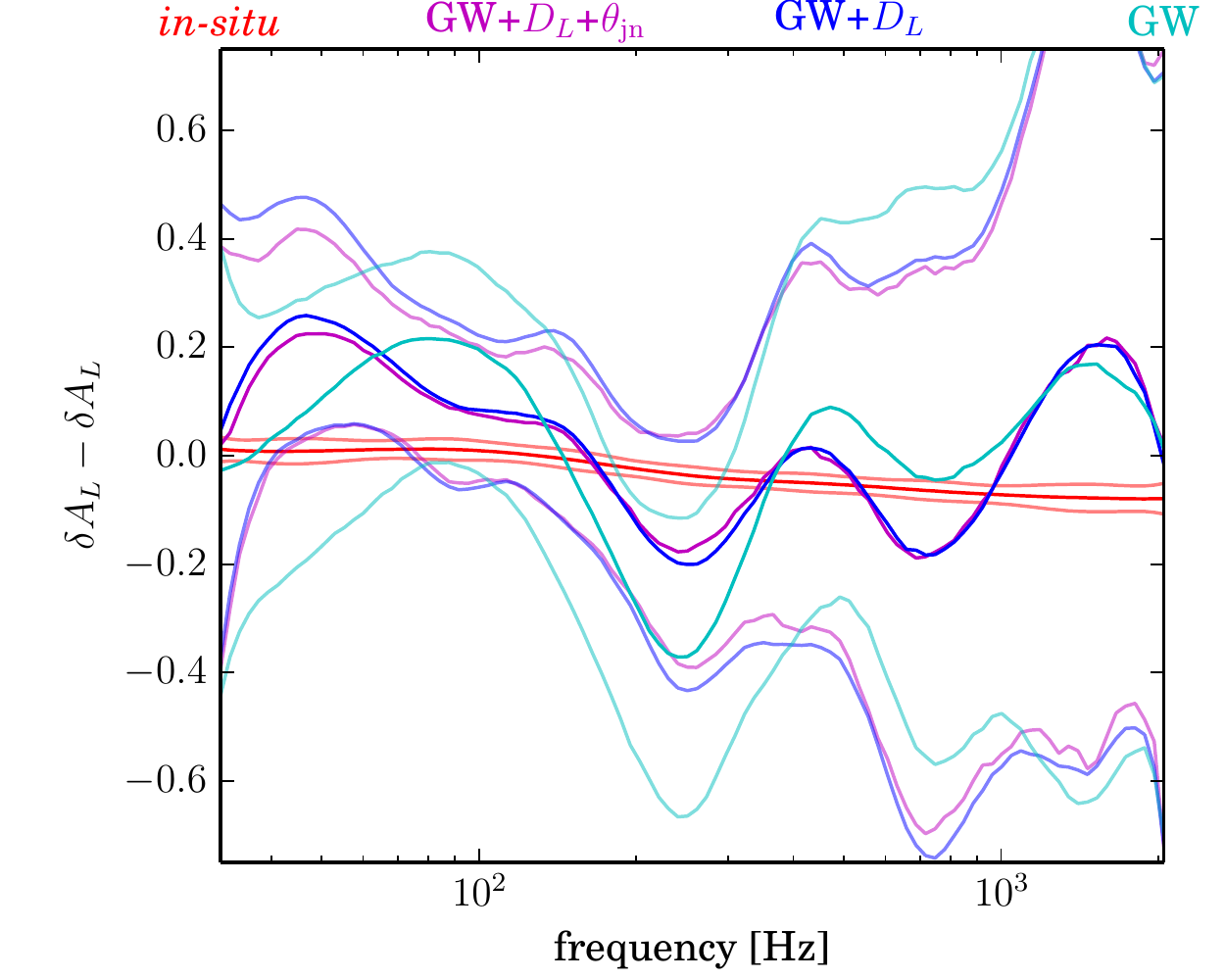}
    \includegraphics[width=1.0\columnwidth, clip=True, trim=0.65cm 0.05cm 0.28cm 0.50cm]{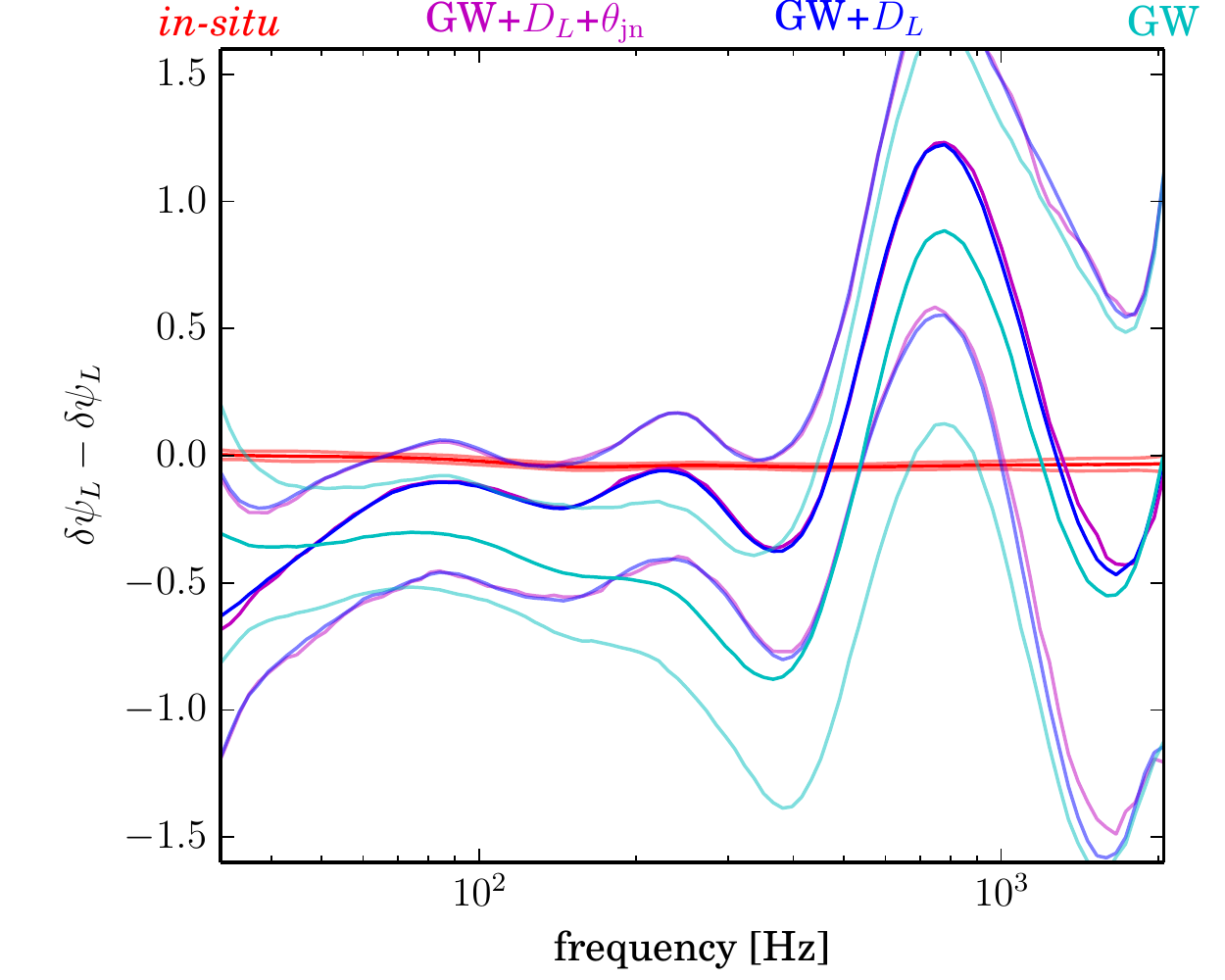}
    \caption{
        Relative calibration between the Hanford and Livingston detectors; color schemes and naming conventions are identical to Figure~\ref{figure:processes}.
        We note that, unlike in Figure~\ref{figure:processes}, additional constraints on $D_L$ and $\theta_\mathrm{jn}$ only marginally impact our ability to measure the relative calibration between detectors.
    }
    \label{figure:relative processes}
\end{figure}

Figures~\ref{figure:processes} and~\ref{figure:relative processes} show the constraints on amplitude and phase calibration in the two LIGO detectors from GW170817.\footnote{While Virgo data was used in this analysis, GW170817 does not constrain Virgo's calibration due to the low signal-to-noise ratio in that detector.}
We show several posteriors
\begin{itemize}
    \item{(GW): GW data alone with loose calibration priors. We assume standard (uninformative) priors for all intrinsic and extrinsic parameters. In particular, we assume sources are uniformly distributed in volume ignoring local cosmological effects: $p(D_L) \propto D_L^2$.}
    \item{(GW+$D_L$): loose calibration priors and the source's location constrained by the approximate uncertainty in NGC 4993's location. We assume conservative bounds of $D_L\in[32,\,50]$\,Mpc with $p(D_L)\propto D_L^2$.}
    \item{(GW+$D_L$+$\theta_\mathrm{jn}$): loose calibration priors using the EM constraint on the source location as well as approximate constraints on the source's inclination from EM observations. We again use somewhat conservative constraints of $\theta_\mathrm{jn}\in [10^\circ,\,30^\circ]\oplus[150^\circ,170^\circ]$ while~\cite{Mooley2018} suggests $14^\circ \leq \theta_\mathrm{jn} \leq 28^\circ$ (see also~\cite{2017ApJ...851L..36G,Mooley2018}).}
    \item{(\textit{in-situ}): GW data alone but with the directly-measured calibration prior uncertainties~\cite{Catalog, cal_envelop}.}
\end{itemize}

These figures elucidate several important points.
We find that while EM constraints impact the amplitude uncertainty, $\delta A$, all posterior processes for $\delta A$ assuming loose calibration priors are significantly larger than the directly-measured calibration uncertainties.
In addition, the EM constraints do not meaningfully affect the posterior uncertainty estimation of $\delta \psi$.
However, EM constraints do affect the inferred source-frame chirp mass; this is because it depends on both the detector frame chirp mass and $D_{L}$.
In this sense, we find astronomical calibration is, in practice, only a corroboration of \textit{in-situ}, directly measured calibration.

Because all detectors observe the same signal, uncertainties in the parameters of that signal manifest as correlations between neighboring frequencies in the calibration model, both within a single detector and between detectors.
We do not observe measurable correlations between the amplitude and phase uncertainties, although each separately correlates with a few signal parameters.

We first focus on the amplitude calibration. $\delta A$ primarily correlates with $D_L$ and $\theta_\mathrm{jn}$ because the amplitude observed by each detector is a combination of all three (Eqs.~\ref{model}, \ref{plus_strain}, and \ref{cross_strain}).
$\delta A$ tends to move coherently at all frequencies, with larger $\delta A$ corresponding to larger $D_L$ or smaller $\cos\theta_{jn}$.
This degeneracy means that GW data alone constrains only the relative (frequency-dependent) calibration; constraints on the absolute (frequency-independent) calibration remains dominated by our prior.
We note that precise knowledge of either $D_L$ or $\theta_\mathrm{jn}$ is insufficient to improve the posterior constraints on $\delta A$.
Instead, we only see significantly tighter $\delta A$ constraints when we include knowledge of both $D_L$ and $\theta_\mathrm{jn}$ from EM observations.
This is because GW170817 was primarily witnessed by the two LIGO detectors, and they are effectively sensitive to only a single polarization because they are nearly aligned.
This means that the LIGOs alone cannot constrain the inclination significantly, and the uncertainty in $\theta_\mathrm{jn}$ is large enough that, even with relatively precise knowledge of $D_L$, we still observe large posterior uncertainty for $\delta A$.
Similarly, knowledge of $\theta_\mathrm{jn}$ without knowledge of $D_L$ means the uncertainty in $D_L$ limits the precision of our inference of $\delta A$.
More favorable source locations, in which more than two detectors witness the signal with large signal-to-noise ratios, may allow the GW data alone to constrain $\theta_\mathrm{jn}$.
In this case, we would only rely on EM measurements of $D_L$ (alternatively the redshift, assuming $H_0$ is well known), which likely have smaller systematic errors than the associated constraints on $\theta_\mathrm{jn}$ from GRB and kilonova modeling (see, for example, \cite{Fernandez2015, Nakar2018, Alexander2018, Kasen2017, Kasen2014}).

\begin{figure*}
    \includegraphics[width=0.95\textwidth, clip=True, trim=0.25cm 0.25cm 0.50cm 0.50cm]{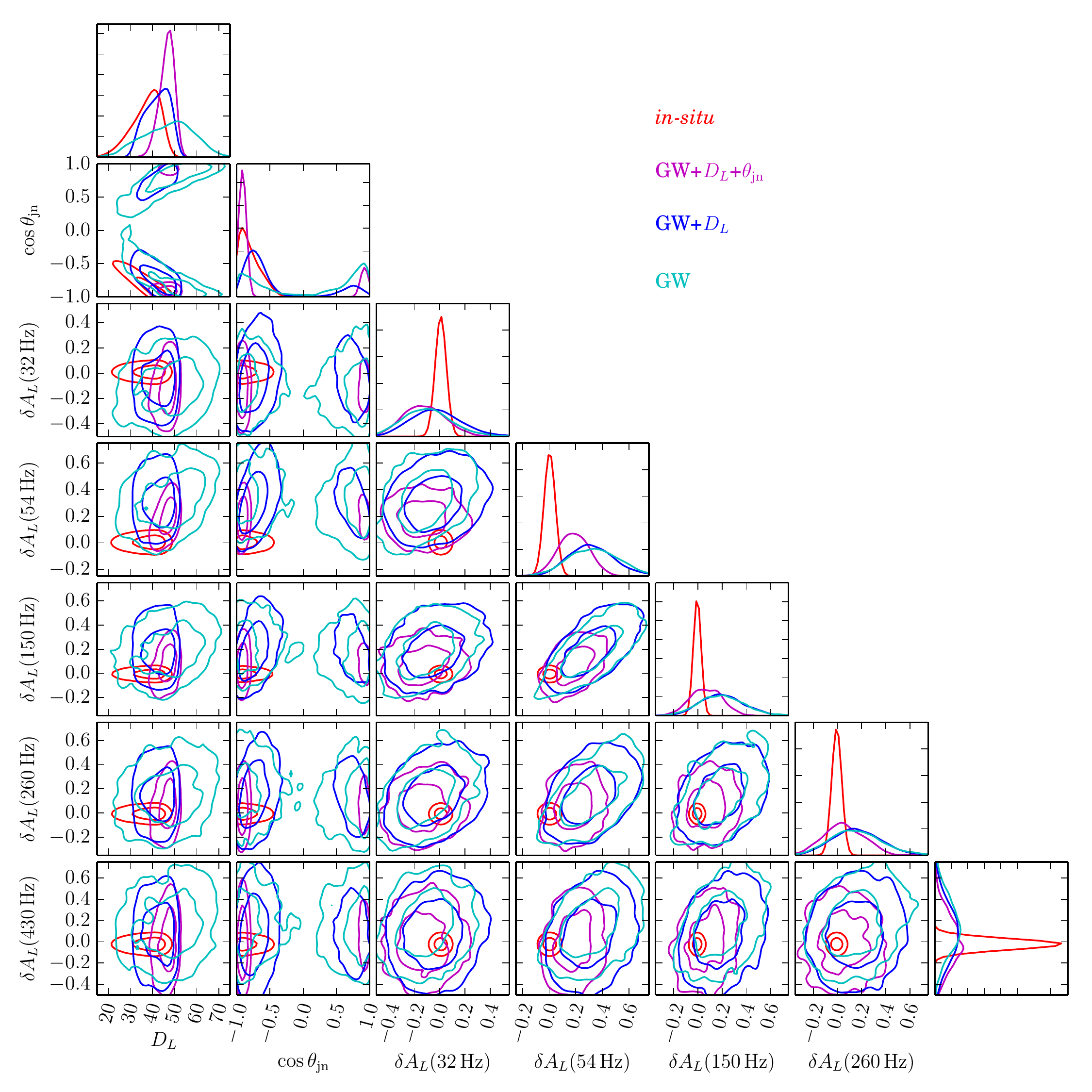}
    \caption{
        Joint and marginal posterior distributions for $D_L$, $\theta_\mathrm{jn}$, and $\delta A$ at four representative frequencies in the LIGO Livingston detector for the same set of priors as Figure~\ref{figure:processes}.
        Contours represent 50\% and 90\% credible regions.
        We note that all $\delta A$ are correlated, although the correlations' strength decreases as the frequency separation increases.
        We also note that the $\delta A$ constraints are essentially unchanged when we only include EM information for $D_L$.
        Only EM constraints for both $D_L$ and $\theta_\mathrm{jn}$ improve the precision of $\delta A$.
    }
    \label{figure:amp corner}
\end{figure*}

Nonetheless, even with GW data alone, GW170817 provides non-trivial posterior constraints on the relative $\delta A$ between $\sim$40--300 Hz, with roughly $\pm\relampSTD$ precision (see Fig.~\ref{figure:relative processes}).
The precision of the absolute amplitude calibration is similar to the relative precision for GW170817, although the absolute calibration is likely strongly influenced by our prior.
We re-emphasize that the directly-measured calibration priors are significantly tighter, and we do not learn anything about the detector calibration if we use those measurements as priors regardless of which EM constraints are imposed.
Figure~\ref{figure:amp corner} shows the marginal and joint posterior distributions for the amplitude calibration uncertainties at four representative frequencies.
The amplitude calibration posteriors' shapes and the imposed constraints can be understood through the toy model in Figure~\ref{figure:cartoon}.

\begin{figure*}
    \includegraphics[width=0.95\textwidth, clip=True, trim=0.25cm 0.00cm 0.50cm 0.50cm]{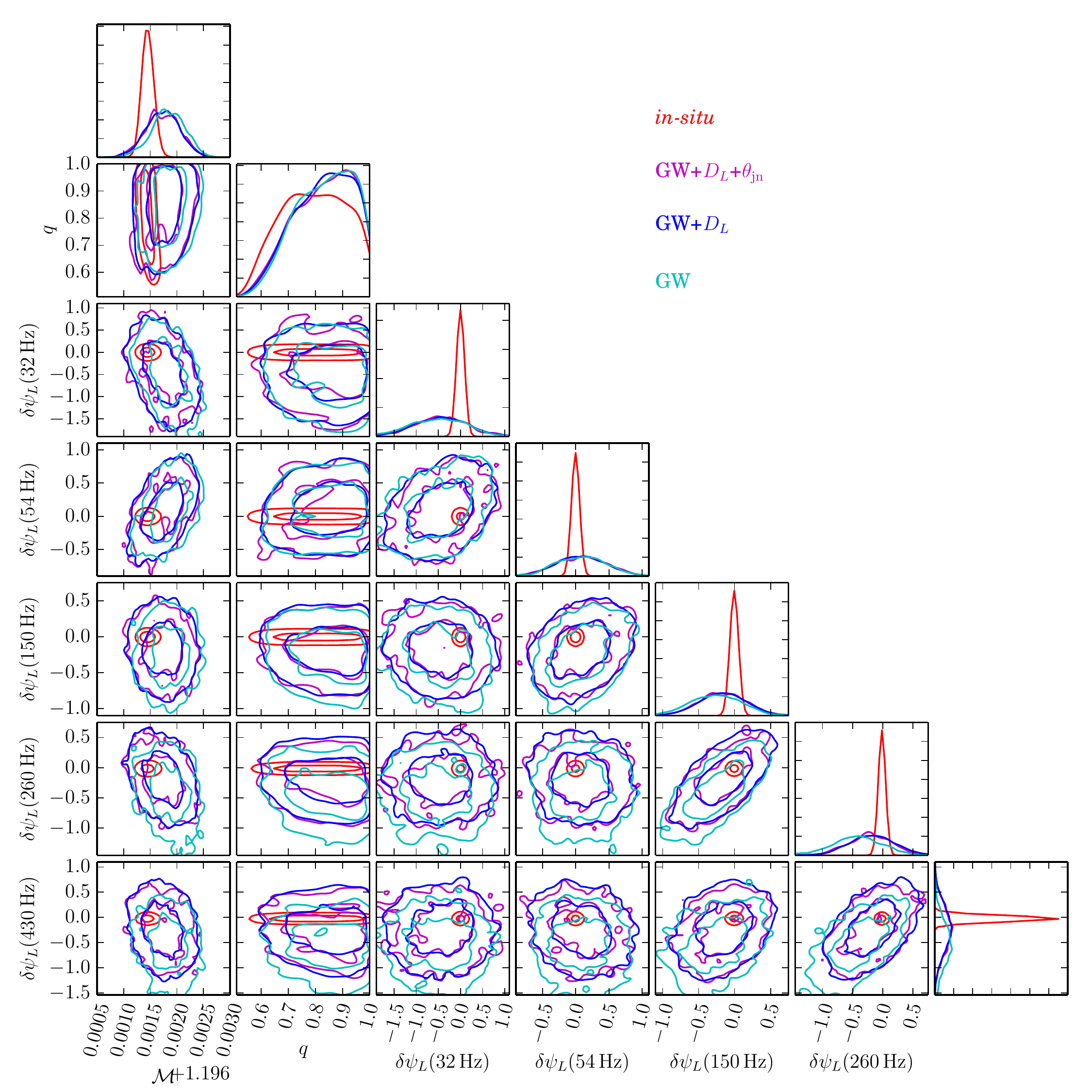}
    \caption{
        Joint and marginal posterior distributions for $\mathcal{M}$, the mass ratio $q=m_2/m_1$, and $\delta \psi$ at a few frequencies in the LIGO Livingston detector for the same set of priors as Figure~\ref{figure:processes}.
        Contours show the 50\% and 90\% credible regions.
        Note that EM information about $D_L$ and $\theta_\mathrm{jn}$ does not affect $\delta \psi$ significantly.
        Furthermore, we observe slight negative correlations between $\delta \psi(54\,\mathrm{Hz})$ and $\delta \psi(430\,\mathrm{Hz})$ and a reversal in the correlations between $\mathcal{M}$ and $\delta\psi$, suggesting the phase errors rotate around the ``best measured frequency'' ($\sim$100 Hz).
    }
    \label{figure:phs corner}
\end{figure*}

We now turn to the phase calibration. While $\delta \psi$ does not correlate with extrinsic parameters or $\delta A$, it does correlate with $\mathcal{M}$.
In contrast to $\delta A$, which tends to move in the same direction at all frequencies, $\delta \psi$ tends to rotate around the ``best measured'' phase at the detectors' most sensitive frequency ($\sim$100 Hz).
This is because $\mathcal{M}$ determines the rate at which the signal's phase changes, and the signal's phase is reasonably well measured at that frequency.
Therefore, if $\mathcal{M}$ changes, the signals phase rotates around the well measured point and $\delta \psi$ changes to match the observed data; note the inverted correlations between $\mathcal{M}$ and $\delta\psi(54\,\mathrm{Hz})$ and $\mathcal{M}$ and $\delta\psi(430\,\mathrm{Hz})$ in Figure~\ref{figure:phs corner}.
While neighboring frequencies are still highly correlated, we additionally observe slight anti-correlations between $\delta \psi(54\,\mathrm{Hz})$ and $\delta \psi(430\,\mathrm{Hz})$ whereas we only see positive correlations between $\delta A(54\,\mathrm{Hz})$ and $\delta A(430\,\mathrm{Hz})$.
The \textit{in-situ} calibration uncertainties are tight, and the likelihood is flat across their entire support.
We do not further constrain $\delta \psi$ with GW170817 when using the official priors.
Again, while the GW data primarily constrains the relative phase calibration, we note that EM observations do not constrain $\phi_o$ and therefore do not constrain the absolute phase further.

We also note that the calibration uncertainties in each detector could be composed of several distinct components.
For example, each LIGO \texttt{PCAL} system relies on photodiodes calibrated against the same NIST-traceable ``gold standard''~\cite{Karki2016}.
Therefore, miscalibration within the gold standard would manifest as calibration systematics shared across both detectors.
We decompose the calibration error in each detector into separate components (e.g. $\delta A_{H,\mathrm{tot}} = \delta A + \delta A_H$, where $\delta A$ is shared between all detectors and the $\delta A_H$, $\delta A_L$, etc. are independent; see Appendix~\ref{section:decomposition} for more details).
Generally, the additional model freedom associated with marginalizing over independent components in each detector to constrain the shared uncertainty inflates the uncertainty relative to purely independent calibration errors,\footnote{$\delta A_\mathrm{tot}$ is composed of the sum of multiple terms and the data only constrains the sum; uncertainty in any one of the terms will correspond to larger marginal uncertainty in the others.} and GW170817 constrains the shared absolute amplitude uncertainty to $\pm\sharedabsampSTD$ and the absolute phase to $\pm\sharedabsphsSTD$ when assuming identical, independent zero-mean Gaussian priors on both the shared and independent components.
Interestingly, the precision of the shared parameters appears to be much less affected by EM constraints imposed on $D_L$ and $\theta_\mathrm{jn}$.

\section{Calibration from future events}
\label{section:scaling}

GW170817 serves as a useful proof-of-principle of an astronomical calibrator. We note that, while it has the largest signal-to-noise ratio ($\rho$) of any GW source detected to date, it still produces constraints more than an order of magnitude worse than the official calibration uncertainties.
Competitive astrophysical calibration will only come from combining information from multiple events---as each event provides marginally more information, the cumulative effect will reduce the overall calibration uncertainty.
In particular, this may identify unknown or unexpected calibration systematics.

To investigate this further, we adopt simple models for how astronomical calibration uncertainties will scale with additional events.
The calibration likelihoods are nearly Gaussian, so we combine multiple events by multiplying Gaussian likelihoods along with a prior.
When we have only GW data, we will only constrain the relative calibration, and the uncertainty is expected to scale as (see Appendix~\ref{section:toy model}):
\begin{equation}
    \sigma_{\delta A_1 - \delta A_2}^2 = \sigma_\mathrm{prior}^2\left(\frac{\sigma_\mathrm{like}^2}{\sigma_\mathrm{like}^2 + 2\sigma_\mathrm{prior}^2}\right),
\end{equation}
where the cumulative likelihood is determined by summing the likelihoods from each event in quadrature:
\begin{equation}\label{likelihood}
    \sigma_\mathrm{like}^{-2} = \sum\limits_{i} \left(\frac{\sigma_o}{\rho_i}\right)^{-2}.
\end{equation}
and $\sigma_\mathrm{prior}$ is the \textit{a priori} uncertainty for each $\delta A$ separately.
The marginal uncertainty on the absolute calibration, on the other hand, should scale as
\begin{equation}
    \sigma_\mathrm{\delta A_1 + \delta A_2}^{2} = \sigma_\mathrm{prior}^2\left(\frac{\sigma_\mathrm{like}^2 + \sigma_\mathrm{prior}^2}{\sigma_\mathrm{like}^2 + 2\sigma_\mathrm{prior}^2}\right).
\end{equation}
In the limit of many events, the relative calibration precision will scale as $\sigma \propto 1/\sqrt{N}$ whereas the marginal absolute calibration will still be prior dominated.

\begin{figure}
    \includegraphics[width=0.49\textwidth, clip=True, trim=0.00cm 0.00cm 1.00cm 0.50cm]{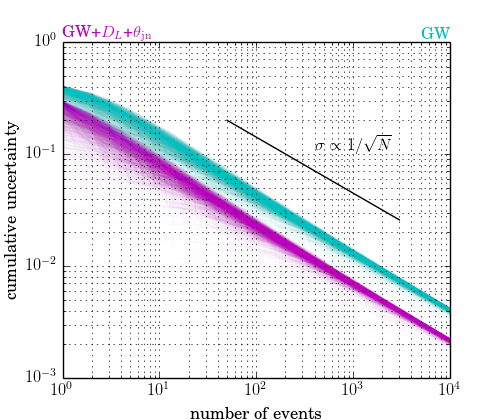}
    \caption{
        Expected scaling of the cumulative amplitude uncertainty with a population of events.
        Each curve shows one realization of our simulation assuming parameters consistent with our wide calibration priors: (\textit{light blue}: GW) relative calibration uncertainty with no EM constraints and (\textit{purple}: GW+$D_L$+$\theta_\mathrm{jn}$) absolute uncertainty with both GW and EM constraints.
    }
    \label{figure:scaling}
\end{figure}

For each event we draw $\rho_i$ from a distribution $\sim \rho^{-4}$~\cite{2014arXiv1409.0522C} and require $\rho_i\geq12$.
We draw many independent events, stack them, and repeat this process for many trials.
Using parameters characteristic of our loose calibration priors and the amplitude constraints obtained from GW170817, Figure~\ref{figure:scaling} shows that we will need only \relnumTEN~events before we constrain the relative amplitude calibration uncertainty to less than 10\%, but \relnumONE~events before we reach 1\%.

If we assume EM constraints comparable to those from GW170817, the nature of the scaling changes.
We expect the joint constraints from both GW and EM data to meaningfully constrain both the relative and absolute amplitude calibration, so the cumulative uncertainty for the amplitude uncertainty would then separately scale as
\begin{equation}
    \sigma_{\delta A}^{-2} = \sigma_\mathrm{prior}^{-2} + \sigma_\mathrm{like}^{-2}.
\end{equation}
In this case, we need \absnumONEem~events to reach 1\% calibration uncertainty.

We note that these cumulative measurements will not be dominated by a single event.
For a single, loud event to constrain $\delta A$ to percent-level uncertainty, it would need to be $\sim$10 times closer than GW170817 assuming comparable EM constraints and similar GW detector sensitivity.
This means the source would need to be within $\sim$4\,Mpc, approximately the radius of the local group.

Following the approach in Section~\ref{section:results}, we decompose the calibration uncertainty into several pieces.
Because the LIGO gold standard photodiodes are only periodically recalibrated, any systematic due to miscalibration of the gold standard would be shared between the LIGOs and could persist for several detections.
As we saw previously, marginalizing over the independent components of the calibration uncertainty typically only increases the uncertainty in the shared parameters, but we expect the same scalings to hold with a larger $\sigma_o$ (Eq.~\ref{likelihood}).
Based on the values found for GW170817, it will take roughly 50\% more events to constrain a calibration parameter shared between detectors while marginalizing over independent components to the same precision as purely independent calibration errors.
This rule of thumb should hold for both relative calibration uncertainty (with or without EM data) and absolute calibration (with EM data).

\section{Discussion}
\label{section:conclusions}

Having established that astrophysical calibration of GW detectors is indeed possible and determining the approximate number of events needed before this method will be competitive with existing calibration schemes, the obvious question remaining is how long it will take to detect that many events.
Based on current rate estimates, the advanced LIGO and Virgo detectors are expected to detect 4--80 BNS and several hundred BBH per year at design sensitivity~\cite{ObservingScenarios}.
While the uncertainties on these detection rates are quite broad, this suggests $\sim6$ years of BNS detections (assuming each has an informative EM counterpart) and nearly a decade of BBH detections (only constraining relative calibration uncertainty assuming GW data alone) before astrophysical calibration can constrain amplitude uncertainty below 1\%.
It is therefore possible, although not altogether likely, that astrophysical calibration could become competitive within the LIGO facilities' lifetimes.
Even at that point, though, it is quite likely that astronomical calibration will only corroborate existing \textit{in-situ} techniques, which may become more precise themselves.

It's worth noting, however, that the gold standard photodiodes used in the LIGO \texttt{PCAL}s are recalibrated approximately once per year~\cite{pcal_nist}.
If astrophysical calibration were to constrain unknown systematics within those photodiodes, we would only be able to use the detections made between photodiode recalibrations.
This implies astrophysical calibration could only ever constrain amplitude uncertainties to somewhere between \limnum.

We note that better models of possible calibration errors, rather than simple spline interpolation, could prove more constraining.
This could also provide informative constraints when the signal is below the noise, such as is the case above $\sim300$ Hz for GW170817.
However, this is not our primary goal.
Instead, the model freedom associated with spline interpolation should allow astrophysical calibration to detect a more general class of calibration errors, not just those described by a particular model.
Indeed, it is possible, although not certain, that astrophysical calibration could be the only way to detect such deviations if they manifest within the calibration of LIGO's gold standard photodiode.

Confirmation binaries expected for the {\it Laser Interferometer Space Antenna} ({\it LISA},~\cite{LISA}) are in many ways analogous to the astronomical calibrators discussed here, and our approach could be used to construct a similar calibration for space-based gravitational-wave detectors.
However, \textit{LISA}'s calibration is expected to be significantly better than that of ground-based detectors, since it is fundamentally limited by the ability to isolate the test masses, and this has already been established by {\it LISA Pathfinder}\/ to work well past the 1\% level~\cite{LISAPathfinder,PhysRevLett.120.061101}.

Finally, we turn our attention to the implications for $H_0$ measurements with GW standard sirens.
GW170817 and AT 2017gfo produced the first such measurement: $H_0 = 70^{+12}_{-8}$ km s$^{-1}$ Mpc$^{-1}$~\cite{H0}.
While the precision is remarkable given it is based on a single event, it is not yet precise enough to resolve the percent-level discrepancy between the Planck~\cite{Planck} and SHoES~\cite{SHoES} results.
Similarly, recent work constrains $H_0$ using a statistical approach in the absence of EM counterparts~\cite{Fishbach2018}.
Typically, the statistical approach will provide weaker constraints than events with EM counterparts~\cite{Chen2017}.
However, the precision of both methods will be fundamentally limited by our ability to precisely calibrate GW detectors.

Unfortunately, we will not be able to simultaneously calibrate the absolute amplitude response of our detectors and infer $H_0$ because $H_0$ is degenerate with $D_L$ (assuming a fixed redshift) and therefore the absolute amplitude uncertainty within our detectors.
Indeed, the EM constraints we applied in our analysis of GW170817 are primarily limited by uncertainty in $H_0$.
This holds for BBH without EM counterparts and BNS with electromagnetic counterparts, as we've shown that knowledge of only $\theta_\mathrm{jn}$ is not enough to constrain the absolute calibration.

Nonetheless, we will be able to constrain the frequency-dependent relative amplitude and phase calibration of our detectors regardless of our prior knowledge of $H_0$.
While it is unlikely astrophysical astrophysical constraints will ever surpass direct \textit{in-situ} measurements, they could provide a meaningful independent corroboration with uncertainties only a factor of a few wider within the advanced detectors' lifetimes.

\acknowledgments

The authors are thankful for many useful conversations with Jeff Kissel and the entire the LIGO Scientific Collaboration Calibration Group.
We also acknowledge discussions with Curt Cutler and Pete Bender about \textit{LISA} calibration.
R.E. and D.E.H. are supported at the University of Chicago by the Kavli Institute for Cosmological Physics through an endowment from the Kavli Foundation and its founder Fred Kavli.
D.E.H. is also supported by NSF grant PHY-1708081, and gratefully acknowledges the Marion and Stuart Rice Award.
The authors also gratefully acknowledge the computational resources provided by the LIGO Laboratory and supported by NSF grants PHY-0757058 and PHY-0823459.

\bibliography{refs}

\begin{thebibliography}{47}%
\makeatletter
\providecommand \@ifxundefined [1]{%
 \@ifx{#1\undefined}
}%
\providecommand \@ifnum [1]{%
 \ifnum #1\expandafter \@firstoftwo
 \else \expandafter \@secondoftwo
 \fi
}%
\providecommand \@ifx [1]{%
 \ifx #1\expandafter \@firstoftwo
 \else \expandafter \@secondoftwo
 \fi
}%
\providecommand \natexlab [1]{#1}%
\providecommand \enquote  [1]{``#1''}%
\providecommand \bibnamefont  [1]{#1}%
\providecommand \bibfnamefont [1]{#1}%
\providecommand \citenamefont [1]{#1}%
\providecommand \href@noop [0]{\@secondoftwo}%
\providecommand \href [0]{\begingroup \@sanitize@url \@href}%
\providecommand \@href[1]{\@@startlink{#1}\@@href}%
\providecommand \@@href[1]{\endgroup#1\@@endlink}%
\providecommand \@sanitize@url [0]{\catcode `\\12\catcode `\$12\catcode
  `\&12\catcode `\#12\catcode `\^12\catcode `\_12\catcode `\%12\relax}%
\providecommand \@@startlink[1]{}%
\providecommand \@@endlink[0]{}%
\providecommand \url  [0]{\begingroup\@sanitize@url \@url }%
\providecommand \@url [1]{\endgroup\@href {#1}{\urlprefix }}%
\providecommand \urlprefix  [0]{URL }%
\providecommand \Eprint [0]{\href }%
\providecommand \doibase [0]{http://dx.doi.org/}%
\providecommand \selectlanguage [0]{\@gobble}%
\providecommand \bibinfo  [0]{\@secondoftwo}%
\providecommand \bibfield  [0]{\@secondoftwo}%
\providecommand \translation [1]{[#1]}%
\providecommand \BibitemOpen [0]{}%
\providecommand \bibitemStop [0]{}%
\providecommand \bibitemNoStop [0]{.\EOS\space}%
\providecommand \EOS [0]{\spacefactor3000\relax}%
\providecommand \BibitemShut  [1]{\csname bibitem#1\endcsname}%
\let\auto@bib@innerbib\@empty
\bibitem [{\citenamefont {Abbott}\ \emph
  {et~al.}(2017{\natexlab{a}})\citenamefont {Abbott} \emph
  {et~al.}}]{GW170817}%
  \BibitemOpen
  \bibfield  {author} {\bibinfo {author} {\bibfnamefont {B.~P.}\ \bibnamefont
  {Abbott}} \emph {et~al.} (\bibinfo {collaboration} {LIGO Scientific
  Collaboration and Virgo Collaboration}),\ }\href {\doibase
  10.1103/PhysRevLett.119.161101} {\bibfield  {journal} {\bibinfo  {journal}
  {Phys. Rev. Lett.}\ }\textbf {\bibinfo {volume} {119}},\ \bibinfo {pages}
  {161101} (\bibinfo {year} {2017}{\natexlab{a}})}\BibitemShut {NoStop}%
\bibitem [{\citenamefont {Abbott}\ \emph {et~al.}(2019)\citenamefont {Abbott}
  \emph {et~al.}}]{SourceProperties}%
  \BibitemOpen
  \bibfield  {author} {\bibinfo {author} {\bibfnamefont {B.~P.}\ \bibnamefont
  {Abbott}} \emph {et~al.} (\bibinfo {collaboration} {LIGO Scientific
  Collaboration and Virgo Collaboration}),\ }\href {\doibase
  10.1103/PhysRevX.9.011001} {\bibfield  {journal} {\bibinfo  {journal} {Phys.
  Rev. X}\ }\textbf {\bibinfo {volume} {9}},\ \bibinfo {pages} {011001}
  (\bibinfo {year} {2019})}\BibitemShut {NoStop}%
\bibitem [{\citenamefont {Abbott}\ \emph
  {et~al.}(2018{\natexlab{a}})\citenamefont {Abbott} \emph {et~al.}}]{EOS}%
  \BibitemOpen
  \bibfield  {author} {\bibinfo {author} {\bibfnamefont {B.~P.}\ \bibnamefont
  {Abbott}} \emph {et~al.} (\bibinfo {collaboration} {The LIGO Scientific
  Collaboration and the Virgo Collaboration}),\ }\href {\doibase
  10.1103/PhysRevLett.121.161101} {\bibfield  {journal} {\bibinfo  {journal}
  {Phys. Rev. Lett.}\ }\textbf {\bibinfo {volume} {121}},\ \bibinfo {pages}
  {161101} (\bibinfo {year} {2018}{\natexlab{a}})}\BibitemShut {NoStop}%
\bibitem [{\citenamefont {Landry}\ and\ \citenamefont
  {Essick}(2018)}]{Landry2018}%
  \BibitemOpen
  \bibfield  {author} {\bibinfo {author} {\bibfnamefont {P.}~\bibnamefont
  {Landry}}\ and\ \bibinfo {author} {\bibfnamefont {R.}~\bibnamefont
  {Essick}},\ }\href@noop {} {\  (\bibinfo {year} {2018})},\ \Eprint
  {http://arxiv.org/abs/1811.12529} {arXiv:1811.12529 [gr-qc]} \BibitemShut
  {NoStop}%
\bibitem [{\citenamefont {Abbott}\ \emph
  {et~al.}(2017{\natexlab{b}})\citenamefont {Abbott} \emph {et~al.}}]{GRB}%
  \BibitemOpen
  \bibfield  {author} {\bibinfo {author} {\bibfnamefont {B.~P.}\ \bibnamefont
  {Abbott}} \emph {et~al.},\ }\href
  {http://stacks.iop.org/2041-8205/848/i=2/a=L13} {\bibfield  {journal}
  {\bibinfo  {journal} {The Astrophysical Journal Letters}\ }\textbf {\bibinfo
  {volume} {848}},\ \bibinfo {pages} {L13} (\bibinfo {year}
  {2017}{\natexlab{b}})}\BibitemShut {NoStop}%
\bibitem [{\citenamefont {{The LIGO Scientific Collaboration}}\ \emph
  {et~al.}(2017{\natexlab{a}})\citenamefont {{The LIGO Scientific
  Collaboration}}, \citenamefont {{The Virgo Collaboration}} \emph
  {et~al.}}]{multimessenger}%
  \BibitemOpen
  \bibfield  {author} {\bibinfo {author} {\bibnamefont {{The LIGO Scientific
  Collaboration}}}, \bibinfo {author} {\bibnamefont {{The Virgo
  Collaboration}}},  \emph {et~al.},\ }\href
  {http://stacks.iop.org/2041-8205/848/i=2/a=L12} {\bibfield  {journal}
  {\bibinfo  {journal} {The Astrophysical Journal Letters}\ }\textbf {\bibinfo
  {volume} {848}},\ \bibinfo {pages} {L12} (\bibinfo {year}
  {2017}{\natexlab{a}})}\BibitemShut {NoStop}%
\bibitem [{\citenamefont {{The LIGO Scientific Collaboration}}\ and\
  \citenamefont {{The Virgo Collaboration}}(2017)}]{kilonova}%
  \BibitemOpen
  \bibfield  {author} {\bibinfo {author} {\bibnamefont {{The LIGO Scientific
  Collaboration}}}\ and\ \bibinfo {author} {\bibnamefont {{The Virgo
  Collaboration}}},\ }\href {http://stacks.iop.org/2041-8205/850/i=2/a=L39}
  {\bibfield  {journal} {\bibinfo  {journal} {The Astrophysical Journal
  Letters}\ }\textbf {\bibinfo {volume} {850}},\ \bibinfo {pages} {L39}
  (\bibinfo {year} {2017})}\BibitemShut {NoStop}%
\bibitem [{\citenamefont {Coughlin}\ \emph {et~al.}(2018)\citenamefont
  {Coughlin} \emph {et~al.}}]{Coughlin2018}%
  \BibitemOpen
  \bibfield  {author} {\bibinfo {author} {\bibfnamefont {M.~W.}\ \bibnamefont
  {Coughlin}} \emph {et~al.},\ }\href@noop {} {\  (\bibinfo {year} {2018})},\
  \Eprint {http://arxiv.org/abs/1812.04803} {arXiv:1812.04803 [astro-ph.HE]}
  \BibitemShut {NoStop}%
\bibitem [{\citenamefont {{The LIGO Scientific Collaboration}}\ \emph
  {et~al.}(2017{\natexlab{b}})\citenamefont {{The LIGO Scientific
  Collaboration}}, \citenamefont {{The Virgo Collaboraiton}}, \citenamefont
  {{The 1M2H Collaboration}}, \citenamefont {{The Dark Energy Camera GW-EM
  Collaboration and the DES Collaboration}}, \citenamefont {{The DLT40
  Collaboration}}, \citenamefont {{The Las Cumbres Observatory Collaboration}},
  \citenamefont {{The VINROUGE Collaboration}},\ and\ \citenamefont {{The
  MASTER Collaboration}}}]{H0}%
  \BibitemOpen
  \bibfield  {author} {\bibinfo {author} {\bibnamefont {{The LIGO Scientific
  Collaboration}}}, \bibinfo {author} {\bibnamefont {{The Virgo
  Collaboraiton}}}, \bibinfo {author} {\bibnamefont {{The 1M2H
  Collaboration}}}, \bibinfo {author} {\bibnamefont {{The Dark Energy Camera
  GW-EM Collaboration and the DES Collaboration}}}, \bibinfo {author}
  {\bibnamefont {{The DLT40 Collaboration}}}, \bibinfo {author} {\bibnamefont
  {{The Las Cumbres Observatory Collaboration}}}, \bibinfo {author}
  {\bibnamefont {{The VINROUGE Collaboration}}}, \ and\ \bibinfo {author}
  {\bibnamefont {{The MASTER Collaboration}}},\ }\href
  {http://dx.doi.org/10.1038/nature24471} {\bibfield  {journal} {\bibinfo
  {journal} {Nature}\ }\textbf {\bibinfo {volume} {551}},\ \bibinfo {pages}
  {85} (\bibinfo {year} {2017}{\natexlab{b}})}\BibitemShut {NoStop}%
\bibitem [{\citenamefont {{The LIGO Scientific Collaboration}}(2015)}]{LIGO}%
  \BibitemOpen
  \bibfield  {author} {\bibinfo {author} {\bibnamefont {{The LIGO Scientific
  Collaboration}}},\ }\href {http://stacks.iop.org/0264-9381/32/i=7/a=074001}
  {\bibfield  {journal} {\bibinfo  {journal} {Classical and Quantum Gravity}\
  }\textbf {\bibinfo {volume} {32}},\ \bibinfo {pages} {074001} (\bibinfo
  {year} {2015})}\BibitemShut {NoStop}%
\bibitem [{\citenamefont {Acernese}\ \emph {et~al.}(2015)\citenamefont
  {Acernese} \emph {et~al.}}]{Virgo}%
  \BibitemOpen
  \bibfield  {author} {\bibinfo {author} {\bibfnamefont {F.}~\bibnamefont
  {Acernese}} \emph {et~al.},\ }\href
  {http://stacks.iop.org/0264-9381/32/i=2/a=024001} {\bibfield  {journal}
  {\bibinfo  {journal} {Classical and Quantum Gravity}\ }\textbf {\bibinfo
  {volume} {32}},\ \bibinfo {pages} {024001} (\bibinfo {year}
  {2015})}\BibitemShut {NoStop}%
\bibitem [{\citenamefont {Abbott}\ \emph
  {et~al.}(2018{\natexlab{b}})\citenamefont {Abbott} \emph {et~al.}}]{Catalog}%
  \BibitemOpen
  \bibfield  {author} {\bibinfo {author} {\bibfnamefont {B.~P.}\ \bibnamefont
  {Abbott}} \emph {et~al.},\ }\href@noop {} {\  (\bibinfo {year}
  {2018}{\natexlab{b}})},\ \Eprint {http://arxiv.org/abs/1811.12907}
  {arXiv:1811.12907} \BibitemShut {NoStop}%
\bibitem [{\citenamefont {Abbott}\ \emph
  {et~al.}(2018{\natexlab{c}})\citenamefont {Abbott} \emph
  {et~al.}}]{Populations}%
  \BibitemOpen
  \bibfield  {author} {\bibinfo {author} {\bibfnamefont {B.~P.}\ \bibnamefont
  {Abbott}} \emph {et~al.},\ }\href@noop {} {\  (\bibinfo {year}
  {2018}{\natexlab{c}})},\ \Eprint {http://arxiv.org/abs/1811.12940}
  {arXiv:1811.12940} \BibitemShut {NoStop}%
\bibitem [{\citenamefont {Abbott}\ \emph
  {et~al.}(2017{\natexlab{c}})\citenamefont {Abbott} \emph
  {et~al.}}]{PhysRevD.95.062003}%
  \BibitemOpen
  \bibfield  {author} {\bibinfo {author} {\bibfnamefont {B.~P.}\ \bibnamefont
  {Abbott}} \emph {et~al.} (\bibinfo {collaboration} {LIGO Scientific
  Collaboration}),\ }\href {\doibase 10.1103/PhysRevD.95.062003} {\bibfield
  {journal} {\bibinfo  {journal} {Phys. Rev. D}\ }\textbf {\bibinfo {volume}
  {95}},\ \bibinfo {pages} {062003} (\bibinfo {year}
  {2017}{\natexlab{c}})}\BibitemShut {NoStop}%
\bibitem [{\citenamefont {Cahillane}\ \emph {et~al.}(2017)\citenamefont
  {Cahillane} \emph {et~al.}}]{PhysRevD.96.102001}%
  \BibitemOpen
  \bibfield  {author} {\bibinfo {author} {\bibfnamefont {C.}~\bibnamefont
  {Cahillane}} \emph {et~al.},\ }\href {\doibase 10.1103/PhysRevD.96.102001}
  {\bibfield  {journal} {\bibinfo  {journal} {Phys. Rev. D}\ }\textbf {\bibinfo
  {volume} {96}},\ \bibinfo {pages} {102001} (\bibinfo {year}
  {2017})}\BibitemShut {NoStop}%
\bibitem [{\citenamefont {Acernese}\ \emph {et~al.}(2018)\citenamefont
  {Acernese} \emph {et~al.}}]{0264-9381-35-20-205004}%
  \BibitemOpen
  \bibfield  {author} {\bibinfo {author} {\bibfnamefont {F.}~\bibnamefont
  {Acernese}} \emph {et~al.},\ }\href
  {http://stacks.iop.org/0264-9381/35/i=20/a=205004} {\bibfield  {journal}
  {\bibinfo  {journal} {Classical and Quantum Gravity}\ }\textbf {\bibinfo
  {volume} {35}},\ \bibinfo {pages} {205004} (\bibinfo {year}
  {2018})}\BibitemShut {NoStop}%
\bibitem [{\citenamefont {Abadie}\ \emph {et~al.}(2010)\citenamefont {Abadie}
  \emph {et~al.}}]{ABADIE2010223}%
  \BibitemOpen
  \bibfield  {author} {\bibinfo {author} {\bibfnamefont {J.}~\bibnamefont
  {Abadie}} \emph {et~al.},\ }\href {\doibase
  https://doi.org/10.1016/j.nima.2010.07.089} {\bibfield  {journal} {\bibinfo
  {journal} {Nuclear Instruments and Methods in Physics Research Section A:
  Accelerators, Spectrometers, Detectors and Associated Equipment}\ }\textbf
  {\bibinfo {volume} {624}},\ \bibinfo {pages} {223 } (\bibinfo {year}
  {2010})}\BibitemShut {NoStop}%
\bibitem [{\citenamefont {Karki}\ \emph {et~al.}(2016)\citenamefont {Karki}
  \emph {et~al.}}]{Karki2016}%
  \BibitemOpen
  \bibfield  {author} {\bibinfo {author} {\bibfnamefont {S.}~\bibnamefont
  {Karki}} \emph {et~al.},\ }\href {\doibase 10.1063/1.4967303} {\bibfield
  {journal} {\bibinfo  {journal} {Review of Scientific Instruments}\ }\textbf
  {\bibinfo {volume} {87}},\ \bibinfo {pages} {114503} (\bibinfo {year}
  {2016})},\ \Eprint {http://arxiv.org/abs/https://doi.org/10.1063/1.4967303}
  {https://doi.org/10.1063/1.4967303} \BibitemShut {NoStop}%
\bibitem [{\citenamefont {Goetz}\ and\ \citenamefont
  {Savage}(2010)}]{Goetz_2010}%
  \BibitemOpen
  \bibfield  {author} {\bibinfo {author} {\bibfnamefont {E.}~\bibnamefont
  {Goetz}}\ and\ \bibinfo {author} {\bibfnamefont {R.~L.}\ \bibnamefont
  {Savage}},\ }\href {\doibase 10.1088/0264-9381/27/21/215001} {\bibfield
  {journal} {\bibinfo  {journal} {Classical and Quantum Gravity}\ }\textbf
  {\bibinfo {volume} {27}},\ \bibinfo {pages} {215001} (\bibinfo {year}
  {2010})}\BibitemShut {NoStop}%
\bibitem [{\citenamefont {Estevez}\ \emph {et~al.}(2018)\citenamefont
  {Estevez}, \citenamefont {Lieunard}, \citenamefont {Marion}, \citenamefont
  {Mours}, \citenamefont {Rolland},\ and\ \citenamefont
  {Verkindt}}]{2018arXiv180606572E}%
  \BibitemOpen
  \bibfield  {author} {\bibinfo {author} {\bibfnamefont {D.}~\bibnamefont
  {Estevez}}, \bibinfo {author} {\bibfnamefont {B.}~\bibnamefont {Lieunard}},
  \bibinfo {author} {\bibfnamefont {F.}~\bibnamefont {Marion}}, \bibinfo
  {author} {\bibfnamefont {B.}~\bibnamefont {Mours}}, \bibinfo {author}
  {\bibfnamefont {L.}~\bibnamefont {Rolland}}, \ and\ \bibinfo {author}
  {\bibfnamefont {D.}~\bibnamefont {Verkindt}},\ }\href {\doibase
  10.1088/1361-6382/aae95f} {\bibfield  {journal} {\bibinfo  {journal}
  {Classical and Quantum Gravity}\ }\textbf {\bibinfo {volume} {35}},\ \bibinfo
  {pages} {235009} (\bibinfo {year} {2018})}\BibitemShut {NoStop}%
\bibitem [{\citenamefont {Weinberg}\ \emph {et~al.}(2018)\citenamefont
  {Weinberg}, \citenamefont {{The LIGO Scientific Collaboration}},\ and\
  \citenamefont {{The Virgo Collaboration}}}]{NLTides}%
  \BibitemOpen
  \bibfield  {author} {\bibinfo {author} {\bibfnamefont {N.~N.}\ \bibnamefont
  {Weinberg}}, \bibinfo {author} {\bibnamefont {{The LIGO Scientific
  Collaboration}}}, \ and\ \bibinfo {author} {\bibnamefont {{The Virgo
  Collaboration}}},\ }\href@noop {} {\  (\bibinfo {year} {2018})},\ \Eprint
  {http://arxiv.org/abs/1808.08676} {arXiv:1808.08676} \BibitemShut {NoStop}%
\bibitem [{\citenamefont {Pitkin}\ \emph {et~al.}(2016)\citenamefont {Pitkin},
  \citenamefont {Messenger},\ and\ \citenamefont {Wright}}]{Pitkin2016}%
  \BibitemOpen
  \bibfield  {author} {\bibinfo {author} {\bibfnamefont {M.}~\bibnamefont
  {Pitkin}}, \bibinfo {author} {\bibfnamefont {C.}~\bibnamefont {Messenger}}, \
  and\ \bibinfo {author} {\bibfnamefont {L.}~\bibnamefont {Wright}},\ }\href
  {\doibase 10.1103/PhysRevD.93.062002} {\bibfield  {journal} {\bibinfo
  {journal} {Phys. Rev. D}\ }\textbf {\bibinfo {volume} {93}},\ \bibinfo
  {pages} {062002} (\bibinfo {year} {2016})}\BibitemShut {NoStop}%
\bibitem [{\citenamefont {{The LIGO Scientific
  Collaboration}}(2017)}]{CosmicExplorer}%
  \BibitemOpen
  \bibfield  {author} {\bibinfo {author} {\bibnamefont {{The LIGO Scientific
  Collaboration}}},\ }\href {http://stacks.iop.org/0264-9381/34/i=4/a=044001}
  {\bibfield  {journal} {\bibinfo  {journal} {Classical and Quantum Gravity}\
  }\textbf {\bibinfo {volume} {34}},\ \bibinfo {pages} {044001} (\bibinfo
  {year} {2017})}\BibitemShut {NoStop}%
\bibitem [{\citenamefont {Hild}\ \emph {et~al.}(2011)\citenamefont {Hild} \emph
  {et~al.}}]{EinsteinTelescope}%
  \BibitemOpen
  \bibfield  {author} {\bibinfo {author} {\bibfnamefont {S.}~\bibnamefont
  {Hild}} \emph {et~al.},\ }\href
  {http://stacks.iop.org/0264-9381/28/i=9/a=094013} {\bibfield  {journal}
  {\bibinfo  {journal} {Classical and Quantum Gravity}\ }\textbf {\bibinfo
  {volume} {28}},\ \bibinfo {pages} {094013} (\bibinfo {year}
  {2011})}\BibitemShut {NoStop}%
\bibitem [{\citenamefont {Essick}\ \emph {et~al.}(2017)\citenamefont {Essick},
  \citenamefont {Vitale},\ and\ \citenamefont
  {Evans}}]{FreqDepAntennaResponse}%
  \BibitemOpen
  \bibfield  {author} {\bibinfo {author} {\bibfnamefont {R.}~\bibnamefont
  {Essick}}, \bibinfo {author} {\bibfnamefont {S.}~\bibnamefont {Vitale}}, \
  and\ \bibinfo {author} {\bibfnamefont {M.}~\bibnamefont {Evans}},\ }\href
  {\doibase 10.1103/PhysRevD.96.084004} {\bibfield  {journal} {\bibinfo
  {journal} {Phys. Rev. D}\ }\textbf {\bibinfo {volume} {96}},\ \bibinfo
  {pages} {084004} (\bibinfo {year} {2017})}\BibitemShut {NoStop}%
\bibitem [{\citenamefont {Veitch}\ \emph {et~al.}(2015)\citenamefont {Veitch}
  \emph {et~al.}}]{LALInference}%
  \BibitemOpen
  \bibfield  {author} {\bibinfo {author} {\bibfnamefont {J.}~\bibnamefont
  {Veitch}} \emph {et~al.},\ }\href {\doibase 10.1103/PhysRevD.91.042003}
  {\bibfield  {journal} {\bibinfo  {journal} {Phys. Rev. D}\ }\textbf {\bibinfo
  {volume} {91}},\ \bibinfo {pages} {042003} (\bibinfo {year}
  {2015})}\BibitemShut {NoStop}%
\bibitem [{\citenamefont {{LIGO Scientific Collaboration}}(2018)}]{lalsuite}%
  \BibitemOpen
  \bibfield  {author} {\bibinfo {author} {\bibnamefont {{LIGO Scientific
  Collaboration}}},\ }\href {\doibase 10.7935/GT1W-FZ16} {\enquote {\bibinfo
  {title} {{LIGO} {A}lgorithm {L}ibrary - {LALS}uite},}\ }\bibinfo
  {howpublished} {free software (GPL)} (\bibinfo {year} {2018})\BibitemShut
  {NoStop}%
\bibitem [{\citenamefont {Pankow}\ \emph {et~al.}(2017)\citenamefont {Pankow}
  \emph {et~al.}}]{Pankow2017}%
  \BibitemOpen
  \bibfield  {author} {\bibinfo {author} {\bibfnamefont {C.}~\bibnamefont
  {Pankow}} \emph {et~al.},\ }\href
  {http://stacks.iop.org/0004-637X/834/i=2/a=154} {\bibfield  {journal}
  {\bibinfo  {journal} {The Astrophysical Journal}\ }\textbf {\bibinfo {volume}
  {834}},\ \bibinfo {pages} {154} (\bibinfo {year} {2017})}\BibitemShut
  {NoStop}%
\bibitem [{\citenamefont {Buonanno}\ \emph {et~al.}(2009)\citenamefont
  {Buonanno} \emph {et~al.}}]{Buonanno2009}%
  \BibitemOpen
  \bibfield  {author} {\bibinfo {author} {\bibfnamefont {A.}~\bibnamefont
  {Buonanno}} \emph {et~al.},\ }\href {\doibase 10.1103/PhysRevD.80.084043}
  {\bibfield  {journal} {\bibinfo  {journal} {Phys. Rev. D}\ }\textbf {\bibinfo
  {volume} {80}},\ \bibinfo {pages} {084043} (\bibinfo {year}
  {2009})}\BibitemShut {NoStop}%
\bibitem [{\citenamefont {Mooley}\ \emph {et~al.}(2018)\citenamefont {Mooley},
  \citenamefont {Deller}, \citenamefont {Gottlieb}, \citenamefont {Nakar},
  \citenamefont {Hallinan}, \citenamefont {Bourke}, \citenamefont {Frail},
  \citenamefont {Horesh}, \citenamefont {Corsi},\ and\ \citenamefont
  {Hotokezaka}}]{Mooley2018}%
  \BibitemOpen
  \bibfield  {author} {\bibinfo {author} {\bibfnamefont {K.~P.}\ \bibnamefont
  {Mooley}}, \bibinfo {author} {\bibfnamefont {A.~T.}\ \bibnamefont {Deller}},
  \bibinfo {author} {\bibfnamefont {O.}~\bibnamefont {Gottlieb}}, \bibinfo
  {author} {\bibfnamefont {E.}~\bibnamefont {Nakar}}, \bibinfo {author}
  {\bibfnamefont {G.}~\bibnamefont {Hallinan}}, \bibinfo {author}
  {\bibfnamefont {S.}~\bibnamefont {Bourke}}, \bibinfo {author} {\bibfnamefont
  {D.~A.}\ \bibnamefont {Frail}}, \bibinfo {author} {\bibfnamefont
  {A.}~\bibnamefont {Horesh}}, \bibinfo {author} {\bibfnamefont
  {A.}~\bibnamefont {Corsi}}, \ and\ \bibinfo {author} {\bibfnamefont
  {K.}~\bibnamefont {Hotokezaka}},\ }\href {\doibase 10.1038/s41586-018-0486-3}
  {\bibfield  {journal} {\bibinfo  {journal} {Nature}\ }\textbf {\bibinfo
  {volume} {561}},\ \bibinfo {pages} {355} (\bibinfo {year} {2018})},\ \Eprint
  {http://arxiv.org/abs/1806.09693} {arXiv:1806.09693 [astro-ph.HE]}
  \BibitemShut {NoStop}%
\bibitem [{\citenamefont {{Guidorzi}}\ \emph {et~al.}(2017)\citenamefont
  {{Guidorzi}}, \citenamefont {{Margutti}}, \citenamefont {{Brout}},
  \citenamefont {{Scolnic}}, \citenamefont {{Fong}}, \citenamefont
  {{Alexander}}, \citenamefont {{Cowperthwaite}}, \citenamefont {{Annis}},
  \citenamefont {{Berger}}, \citenamefont {{Blanchard}}, \citenamefont
  {{Chornock}}, \citenamefont {{Coppejans}}, \citenamefont {{Eftekhari}},
  \citenamefont {{Frieman}}, \citenamefont {{Huterer}}, \citenamefont
  {{Nicholl}}, \citenamefont {{Soares-Santos}}, \citenamefont {{Terreran}},
  \citenamefont {{Villar}},\ and\ \citenamefont
  {{Williams}}}]{2017ApJ...851L..36G}%
  \BibitemOpen
  \bibfield  {author} {\bibinfo {author} {\bibfnamefont {C.}~\bibnamefont
  {{Guidorzi}}}, \bibinfo {author} {\bibfnamefont {R.}~\bibnamefont
  {{Margutti}}}, \bibinfo {author} {\bibfnamefont {D.}~\bibnamefont {{Brout}}},
  \bibinfo {author} {\bibfnamefont {D.}~\bibnamefont {{Scolnic}}}, \bibinfo
  {author} {\bibfnamefont {W.}~\bibnamefont {{Fong}}}, \bibinfo {author}
  {\bibfnamefont {K.~D.}\ \bibnamefont {{Alexander}}}, \bibinfo {author}
  {\bibfnamefont {P.~S.}\ \bibnamefont {{Cowperthwaite}}}, \bibinfo {author}
  {\bibfnamefont {J.}~\bibnamefont {{Annis}}}, \bibinfo {author} {\bibfnamefont
  {E.}~\bibnamefont {{Berger}}}, \bibinfo {author} {\bibfnamefont {P.~K.}\
  \bibnamefont {{Blanchard}}}, \bibinfo {author} {\bibfnamefont
  {R.}~\bibnamefont {{Chornock}}}, \bibinfo {author} {\bibfnamefont {D.~L.}\
  \bibnamefont {{Coppejans}}}, \bibinfo {author} {\bibfnamefont
  {T.}~\bibnamefont {{Eftekhari}}}, \bibinfo {author} {\bibfnamefont {J.~A.}\
  \bibnamefont {{Frieman}}}, \bibinfo {author} {\bibfnamefont {D.}~\bibnamefont
  {{Huterer}}}, \bibinfo {author} {\bibfnamefont {M.}~\bibnamefont
  {{Nicholl}}}, \bibinfo {author} {\bibfnamefont {M.}~\bibnamefont
  {{Soares-Santos}}}, \bibinfo {author} {\bibfnamefont {G.}~\bibnamefont
  {{Terreran}}}, \bibinfo {author} {\bibfnamefont {V.~A.}\ \bibnamefont
  {{Villar}}}, \ and\ \bibinfo {author} {\bibfnamefont {P.~K.~G.}\ \bibnamefont
  {{Williams}}},\ }\href {\doibase 10.3847/2041-8213/aaa009} {\bibfield
  {journal} {\bibinfo  {journal} {\apj}\ }\textbf {\bibinfo {volume} {851}},\
  \bibinfo {eid} {L36} (\bibinfo {year} {2017})},\ \Eprint
  {http://arxiv.org/abs/1710.06426} {arXiv:1710.06426 [astro-ph.CO]}
  \BibitemShut {NoStop}%
\bibitem [{\citenamefont {{The LIGO Scientific Collaboration}}\ and\
  \citenamefont {{The Virgo Collaboration}}(2018)}]{cal_envelop}%
  \BibitemOpen
  \bibfield  {author} {\bibinfo {author} {\bibnamefont {{The LIGO Scientific
  Collaboration}}}\ and\ \bibinfo {author} {\bibnamefont {{The Virgo
  Collaboration}}},\ }\href {https://dcc.ligo.org/LIGO-P1900040/public}
  {\enquote {\bibinfo {title} {Calibration uncertainty envelope release for
  gwtc-1},}\ }\bibinfo {howpublished} {LIGO Document Control Center} (\bibinfo
  {year} {2018}),\ \bibinfo {note} {p1900040}\BibitemShut {NoStop}%
\bibitem [{\citenamefont {Fern\'andez}\ and\ \citenamefont
  {Metzger}(2016)}]{Fernandez2015}%
  \BibitemOpen
  \bibfield  {author} {\bibinfo {author} {\bibfnamefont {R.}~\bibnamefont
  {Fern\'andez}}\ and\ \bibinfo {author} {\bibfnamefont {B.~D.}\ \bibnamefont
  {Metzger}},\ }\href {\doibase 10.1146/annurev-nucl-102115-044819} {\bibfield
  {journal} {\bibinfo  {journal} {Ann. Rev. Nucl. Part. Sci.}\ }\textbf
  {\bibinfo {volume} {66}},\ \bibinfo {pages} {23} (\bibinfo {year} {2016})},\
  \Eprint {http://arxiv.org/abs/1512.05435} {arXiv:1512.05435 [astro-ph.HE]}
  \BibitemShut {NoStop}%
\bibitem [{\citenamefont {Nakar}\ and\ \citenamefont
  {Piran}(2018)}]{Nakar2018}%
  \BibitemOpen
  \bibfield  {author} {\bibinfo {author} {\bibfnamefont {E.}~\bibnamefont
  {Nakar}}\ and\ \bibinfo {author} {\bibfnamefont {T.}~\bibnamefont {Piran}},\
  }\href {\doibase 10.1093/mnras/sty952} {\bibfield  {journal} {\bibinfo
  {journal} {Mon. Not. Roy. Astron. Soc.}\ }\textbf {\bibinfo {volume} {478}},\
  \bibinfo {pages} {407} (\bibinfo {year} {2018})},\ \Eprint
  {http://arxiv.org/abs/1801.09712} {arXiv:1801.09712 [astro-ph.HE]}
  \BibitemShut {NoStop}%
\bibitem [{\citenamefont {Alexander}\ \emph {et~al.}(2018)\citenamefont
  {Alexander} \emph {et~al.}}]{Alexander2018}%
  \BibitemOpen
  \bibfield  {author} {\bibinfo {author} {\bibfnamefont {K.~D.}\ \bibnamefont
  {Alexander}} \emph {et~al.},\ }\href {\doibase 10.3847/2041-8213/aad637}
  {\bibfield  {journal} {\bibinfo  {journal} {Astrophys. J.}\ }\textbf
  {\bibinfo {volume} {863}},\ \bibinfo {pages} {L18} (\bibinfo {year}
  {2018})},\ \Eprint {http://arxiv.org/abs/1805.02870} {arXiv:1805.02870
  [astro-ph.HE]} \BibitemShut {NoStop}%
\bibitem [{\citenamefont {Kasen}\ \emph {et~al.}(2017)\citenamefont {Kasen},
  \citenamefont {Metzger}, \citenamefont {Barnes}, \citenamefont {Quataert},\
  and\ \citenamefont {Ramirez-Ruiz}}]{Kasen2017}%
  \BibitemOpen
  \bibfield  {author} {\bibinfo {author} {\bibfnamefont {D.}~\bibnamefont
  {Kasen}}, \bibinfo {author} {\bibfnamefont {B.}~\bibnamefont {Metzger}},
  \bibinfo {author} {\bibfnamefont {J.}~\bibnamefont {Barnes}}, \bibinfo
  {author} {\bibfnamefont {E.}~\bibnamefont {Quataert}}, \ and\ \bibinfo
  {author} {\bibfnamefont {E.}~\bibnamefont {Ramirez-Ruiz}},\ }\href {\doibase
  10.1038/nature24453} {\bibfield  {journal} {\bibinfo  {journal} {Nature}\ }
  (\bibinfo {year} {2017}),\ 10.1038/nature24453},\ \bibinfo {note}
  {[Nature551,80(2017)]},\ \Eprint {http://arxiv.org/abs/1710.05463}
  {arXiv:1710.05463 [astro-ph.HE]} \BibitemShut {NoStop}%
\bibitem [{\citenamefont {Kasen}\ \emph {et~al.}(2015)\citenamefont {Kasen},
  \citenamefont {Fernandez},\ and\ \citenamefont {Metzger}}]{Kasen2014}%
  \BibitemOpen
  \bibfield  {author} {\bibinfo {author} {\bibfnamefont {D.}~\bibnamefont
  {Kasen}}, \bibinfo {author} {\bibfnamefont {R.}~\bibnamefont {Fernandez}}, \
  and\ \bibinfo {author} {\bibfnamefont {B.}~\bibnamefont {Metzger}},\ }\href
  {\doibase 10.1093/mnras/stv721} {\bibfield  {journal} {\bibinfo  {journal}
  {Mon. Not. Roy. Astron. Soc.}\ }\textbf {\bibinfo {volume} {450}},\ \bibinfo
  {pages} {1777} (\bibinfo {year} {2015})},\ \Eprint
  {http://arxiv.org/abs/1411.3726} {arXiv:1411.3726 [astro-ph.HE]} \BibitemShut
  {NoStop}%
\bibitem [{\citenamefont {{Chen}}\ and\ \citenamefont
  {{Holz}}(2014)}]{2014arXiv1409.0522C}%
  \BibitemOpen
  \bibfield  {author} {\bibinfo {author} {\bibfnamefont {H.-Y.}\ \bibnamefont
  {{Chen}}}\ and\ \bibinfo {author} {\bibfnamefont {D.~E.}\ \bibnamefont
  {{Holz}}},\ }\href@noop {} {\bibfield  {journal} {\bibinfo  {journal} {arXiv
  e-prints}\ ,\ \bibinfo {eid} {arXiv:1409.0522}} (\bibinfo {year} {2014})},\
  \Eprint {http://arxiv.org/abs/1409.0522} {arXiv:1409.0522 [gr-qc]}
  \BibitemShut {NoStop}%
\bibitem [{\citenamefont {Abbott}\ \emph {et~al.}(2016)\citenamefont {Abbott}
  \emph {et~al.}}]{ObservingScenarios}%
  \BibitemOpen
  \bibfield  {author} {\bibinfo {author} {\bibfnamefont {B.~P.}\ \bibnamefont
  {Abbott}} \emph {et~al.},\ }\href {\doibase 10.1007/lrr-2016-1} {\bibfield
  {journal} {\bibinfo  {journal} {Living Reviews in Relativity}\ }\textbf
  {\bibinfo {volume} {19}},\ \bibinfo {pages} {1} (\bibinfo {year}
  {2016})}\BibitemShut {NoStop}%
\bibitem [{\citenamefont {{LIGO Photon Calibrator Team}}(2015)}]{pcal_nist}%
  \BibitemOpen
  \bibfield  {author} {\bibinfo {author} {\bibnamefont {{LIGO Photon Calibrator
  Team}}},\ }\href@noop {} {\enquote {\bibinfo {title} {Photon calibrator gold
  standard and checking standard nist calibrations},}\ }\bibinfo {howpublished}
  {LIGO Document Control Center} (\bibinfo {year} {2015}),\ \bibinfo {note}
  {t1500036}\BibitemShut {NoStop}%
\bibitem [{\citenamefont {{Amaro-Seoane}}\ \emph {et~al.}(2017)\citenamefont
  {{Amaro-Seoane}} \emph {et~al.}}]{LISA}%
  \BibitemOpen
  \bibfield  {author} {\bibinfo {author} {\bibfnamefont {P.}~\bibnamefont
  {{Amaro-Seoane}}} \emph {et~al.},\ }\href@noop {} {\bibfield  {journal}
  {\bibinfo  {journal} {arXiv e-prints}\ ,\ \bibinfo {eid} {arXiv:1702.00786}}
  (\bibinfo {year} {2017})},\ \Eprint {http://arxiv.org/abs/1702.00786}
  {arXiv:1702.00786 [astro-ph.IM]} \BibitemShut {NoStop}%
\bibitem [{\citenamefont {{Armano}}\ \emph {et~al.}(2018)\citenamefont
  {{Armano}} \emph {et~al.}}]{LISAPathfinder}%
  \BibitemOpen
  \bibfield  {author} {\bibinfo {author} {\bibfnamefont {M.}~\bibnamefont
  {{Armano}}} \emph {et~al.},\ }\href@noop {} {\bibfield  {journal} {\bibinfo
  {journal} {arXiv e-prints}\ ,\ \bibinfo {eid} {arXiv:1812.05491}} (\bibinfo
  {year} {2018})},\ \Eprint {http://arxiv.org/abs/1812.05491} {arXiv:1812.05491
  [astro-ph.IM]} \BibitemShut {NoStop}%
\bibitem [{\citenamefont {Armano}\ \emph {et~al.}(2018)\citenamefont {Armano}
  \emph {et~al.}}]{PhysRevLett.120.061101}%
  \BibitemOpen
  \bibfield  {author} {\bibinfo {author} {\bibfnamefont {M.}~\bibnamefont
  {Armano}} \emph {et~al.},\ }\href {\doibase 10.1103/PhysRevLett.120.061101}
  {\bibfield  {journal} {\bibinfo  {journal} {Phys. Rev. Lett.}\ }\textbf
  {\bibinfo {volume} {120}},\ \bibinfo {pages} {061101} (\bibinfo {year}
  {2018})}\BibitemShut {NoStop}%
\bibitem [{\citenamefont {{Ade}}\ \emph {et~al.}(2016)\citenamefont {{Ade}}
  \emph {et~al.}}]{Planck}%
  \BibitemOpen
  \bibfield  {author} {\bibinfo {author} {\bibfnamefont {P.~A.~R.}\
  \bibnamefont {{Ade}}} \emph {et~al.},\ }\href {\doibase
  10.1051/0004-6361/201525830} {\bibfield  {journal} {\bibinfo  {journal}
  {aap}\ }\textbf {\bibinfo {volume} {594}},\ \bibinfo {eid} {A13} (\bibinfo
  {year} {2016})},\ \Eprint {http://arxiv.org/abs/1502.01589} {arXiv:1502.01589
  [astro-ph.CO]} \BibitemShut {NoStop}%
\bibitem [{\citenamefont {Riess}\ \emph {et~al.}(2016)\citenamefont {Riess}
  \emph {et~al.}}]{SHoES}%
  \BibitemOpen
  \bibfield  {author} {\bibinfo {author} {\bibfnamefont {A.~G.}\ \bibnamefont
  {Riess}} \emph {et~al.},\ }\href
  {http://stacks.iop.org/0004-637X/826/i=1/a=56} {\bibfield  {journal}
  {\bibinfo  {journal} {The Astrophysical Journal}\ }\textbf {\bibinfo {volume}
  {826}},\ \bibinfo {pages} {56} (\bibinfo {year} {2016})}\BibitemShut
  {NoStop}%
\bibitem [{\citenamefont {Fishbach}\ \emph {et~al.}(2019)\citenamefont
  {Fishbach} \emph {et~al.}}]{Fishbach2018}%
  \BibitemOpen
  \bibfield  {author} {\bibinfo {author} {\bibfnamefont {M.}~\bibnamefont
  {Fishbach}} \emph {et~al.},\ }\href {\doibase 10.3847/2041-8213/aaf96e}
  {\bibfield  {journal} {\bibinfo  {journal} {The Astrophysical Journal}\
  }\textbf {\bibinfo {volume} {871}},\ \bibinfo {pages} {L13} (\bibinfo {year}
  {2019})}\BibitemShut {NoStop}%
\bibitem [{\citenamefont {Chen}\ \emph {et~al.}(2018)\citenamefont {Chen},
  \citenamefont {Fishbach},\ and\ \citenamefont {Holz}}]{Chen2017}%
  \BibitemOpen
  \bibfield  {author} {\bibinfo {author} {\bibfnamefont {H.-Y.}\ \bibnamefont
  {Chen}}, \bibinfo {author} {\bibfnamefont {M.}~\bibnamefont {Fishbach}}, \
  and\ \bibinfo {author} {\bibfnamefont {D.~E.}\ \bibnamefont {Holz}},\ }\href
  {\doibase 10.1038/s41586-018-0606-0} {\bibfield  {journal} {\bibinfo
  {journal} {Nature}\ }\textbf {\bibinfo {volume} {562}},\ \bibinfo {pages}
  {545} (\bibinfo {year} {2018})},\ \Eprint {http://arxiv.org/abs/1712.06531}
  {arXiv:1712.06531 [astro-ph.CO]} \BibitemShut {NoStop}%
\end{thebibliography}%

\onecolumngrid
\appendix

\section{Toy model for scaling}
\label{section:toy model}

As a simple toy model, we consider the priors and likelihoods associated with two amplitude calibration errors.
These could correspond to different frequencies within the same detector or errors at the same frequency in different detectors.
\textit{A priori}, we assume independent Gaussian priors for each parameter separately.
GW data will primarily constrain the relative calibration, or the ratio of calibration errors $(1+\delta A_1)/(1+\delta A_2)$.
This is approximately the difference $\delta A_1 - \delta A_2$ assuming $\delta A \ll 1$.
EM data, on the other hand, will constrain the overall calibration, which is approximately the sum $\delta A_1 + \delta A_2$.
These are shown as shaded regions in Figure~\ref{figure:cartoon}.

We expect the joint posterior for $\delta A_1$ and $\delta A_2$ to behave as
\begin{align}
    \log p \sim & -\frac{1}{2}\left( \frac{(\delta A_1)^2 + (\delta A_2)^2}{\sigma_\mathrm{prior}^2} + \frac{((\delta A_1 - \delta A_2) - \mu_\mathrm{GW})^2}{\sigma_\mathrm{GW}^2} + \frac{((\delta A_1 + \delta A_2) - \mu_\mathrm{EM})^2}{\sigma_\mathrm{EM}^2}\right) \\
              = & -\frac{1}{2}\begin{bmatrix} \delta A_1 - \mu_1 & \delta A_2 - \mu_2\end{bmatrix} C^{-1} \begin{bmatrix} \delta A_1 - \mu_1 \\ \delta A_2 - \mu_2 \end{bmatrix}
\end{align}
where
\begin{gather}
    C^{-1} = \begin{bmatrix} \sigma_\mathrm{prior}^{-2} + \sigma_\mathrm{GW}^{-2} + \sigma_\mathrm{EM}^{-2} & \sigma_\mathrm{EM}^{-2} - \sigma_\mathrm{GW}^{-2} \\ \sigma_\mathrm{EM}^{-2} - \sigma_\mathrm{GW}^{-2} & \sigma_\mathrm{prior}^{-2} + \sigma_\mathrm{GW}^{-2} + \sigma_\mathrm{EM}^{-2} \end{bmatrix} \\
    \begin{bmatrix} \mu_1 \\ \mu_2 \end{bmatrix} = C \begin{bmatrix} \mu_\mathrm{EM} \sigma_\mathrm{EM}^{-2} - \mu_\mathrm{GW} \sigma_\mathrm{GW}^{-2} \\ \mu_\mathrm{EM} \sigma_\mathrm{EM}^{-2} + \mu_\mathrm{GW} \sigma_\mathrm{GW}^{-2} \end{bmatrix}.
\end{gather}
and $\mu$'s and $\sigma$'s denote means and standard deviations, respectively.
Note that $C$ is the covariance matrix of our calibration errors, and from this we can extract several useful scalings.
The marginal posterior uncertainty for each $\delta A_i$ is
\begin{equation}
    \sigma_{\delta A_i}^2 = \frac{\sigma_\mathrm{EM}^2 \sigma_\mathrm{GW}^2 + \sigma_\mathrm{EM}^2 \sigma_\mathrm{prior}^2 + \sigma_\mathrm{GW}^2 \sigma_\mathrm{prior}^2}{\sigma_\mathrm{EM}^2\sigma_\mathrm{GW}^2 + 2\sigma_\mathrm{prior}^2 (\sigma_\mathrm{GW}^2 + \sigma_\mathrm{EM}^2) + 4\sigma_\mathrm{prior}^4}.
\end{equation}
We see that the posterior uncertainty decreases only as both the EM and GW likelihoods tighten.
If, for example, there is no EM data ($\sigma_\mathrm{EM} \gg \sigma_\mathrm{prior},\, \sigma_\mathrm{GW}$), we obtain
\begin{equation}\label{correlated}
    \lim\limits_{\sigma_\mathrm{EM}\rightarrow\infty} \sigma_{\delta A_i}^2 = \sigma_\mathrm{prior}^2\left(\frac{\sigma_\mathrm{GW}^2 + \sigma_\mathrm{prior}^2}{\sigma_\mathrm{GW}^2 + 2\sigma_\mathrm{prior}^2}\right).
\end{equation}
This makes it clear that our \textit{a posteriori} uncertainty on each $\delta A_i$ will be dominated by our prior as $\sigma_\mathrm{GW}\rightarrow0$, although there can be tighter constraints on either the relative or absolute calibration uncertainty, depending on the relative size of $\sigma_\mathrm{GW}$ and $\sigma_\mathrm{EM}$.

We adopt the limit $\sigma_\mathrm{EM}\rightarrow\infty$ when simulating multiple events without EM information, and in that case find the approximate scaling of the relative calibration to be
\begin{equation}
    \sigma_{\delta A_1 - \delta A_2}^2 = \sigma_\mathrm{prior}^2\left(\frac{\sigma_\mathrm{like}^2}{\sigma_\mathrm{like}^2 + 2\sigma_\mathrm{prior}^2}\right).
\end{equation}
However, we assume $\sigma_\mathrm{GW}\sim\sigma_\mathrm{EM}$ for events with both GW and EM constraints, which means the posteriors are roughly independent.
Therefore, the cumulative uncertainty for each calibration amplitude separately from stacking these events should scale as
\begin{equation}
    \sigma_{\delta A_i}^{-2} = \sigma_\mathrm{prior}^{-2} + \sigma_\mathrm{like}^{-2}.
\end{equation}

We note that we used general subscripts throughout this appendix ($\delta A_i$) whereas we use subscripts to specifically call out different detectors in Appendix~\ref{section:decomposition}.
This is because the scaling arguments presented here are in some sense more general. 
The analysis in Appendix~\ref{section:decomposition} specifically focuses on calibration errors shared between detectors, whereas the arguments presented above could apply equally well to calibration errors in separate detectors or to errors in the same detector at different frequencies.


\section{Decomposition of calibration uncertainties into shared and independent parameters}
\label{section:decomposition}

We base our inference on the strain recorded in each interferometer and impose independent Gaussian priors on each calibration parameter in each detector.
If we assume we have two detectors, ($H$, $L$), and the actual calibration errors in each detector are the sum of shared components, ($\delta A$, $\delta \psi$), and an independent component, ($\delta A_H$, $\delta \psi_H$, etc), we obtain
\begin{multline}
    p(\delta A, \delta \psi|d) \propto \int d\delta A_H d\delta A_L d\delta \psi_H d\delta \psi_L\, p(d|\delta A+\delta A_H, \delta A+\delta A_L, \delta \psi+\delta \psi_H, \delta \psi+\delta \psi_L) \\ \times p(\delta A)p(\delta \psi)p(\delta A_H)p(\delta A_L)p(\delta \psi_H)p(\delta \psi_L)
\end{multline}
where we take the priors on each calibration parameter to be independent Gaussian distributions with the same standard deviation at all frequencies.
\texttt{LALInference}, instead, currently supports a single prior on $\delta A+\delta A_H$ rather than separate priors on $\delta A$ and $\delta A_H$.
However, we note that
\begin{equation}\label{equation:this one}
    -\frac{1}{2}\left(\frac{\delta A + \delta A_H}{\sigma_A}\right)^2 = -\frac{1}{2}\left(\frac{\delta A}{\sigma_A}\right)^2 - \frac{1}{2}\left(\frac{\delta A_H}{\sigma_A}\right)^2 - \frac{\delta A \delta A_H}{\sigma_A^2},
\end{equation}
and therefore we can transform the samples obtained from \texttt{LALInference} by imposing a \textit{prior correction} to remove the cross term.
For the simple case shown in Eqn.~\ref{equation:this one}, the term proportional to $S$ is equivalent to $+\delta A \delta A_H/\sigma_A^2$.
Furthermore, because the posteriors with the single prior on $\delta A+\delta A_H$ are nearly Gaussian, we extract the covariance matrix of \texttt{LALInference}'s samples and then marginalize analytically.
We thus find:
\begin{multline}
    \ln p \sim -\frac{1}{2}\begin{bmatrix}\delta A + \delta A_H - \mu_{A_H} & \delta A + \delta A_L - \mu_{A_L} & \delta \psi + \delta \psi_H - \mu_{\psi_H} & \delta \psi + \delta \psi_L - \mu_{\psi_L} \end{bmatrix} C^{-1} \begin{bmatrix} \delta A + \delta A_H - \mu_{A_H} \\ \delta A + \delta A_L - \mu_{A_L} \\ \delta \psi + \delta \psi_H - \mu_{\psi_H} \\ \delta \psi + \delta \psi_L - \mu_{\psi_L} \end{bmatrix} \\
            + \begin{bmatrix} \delta A & \delta A & \delta \psi & \delta \psi \end{bmatrix} S \begin{bmatrix} \delta A_H \\ \delta A_L \\ \delta \psi_H \\ \delta \psi_L \end{bmatrix},
\end{multline}
where the $\mu$'s are the means and $C$ is the covariance matrix of the posterior samples obtained from \texttt{LALInference}.
$S$ is a diagonal matrix representing the prior correction.
Furthermore, introducing
\begin{equation}
    B = \begin{bmatrix} 1_{N\times N} & 1_{N\times N} & 0_{N\times N} & 0_{N\times N} \\ 0_{N\times N} & 0_{N\times N} & 1_{N\times N} & 1_{N\times N}\end{bmatrix},
\end{equation}
where $1_{N\times N}$ and $0_{N \times N}$ are the identity and zero matricies with size $N$ (the number of spline anchor points) so that
\begin{equation}
    \begin{bmatrix} \delta A & \delta \psi \end{bmatrix} B = \begin{bmatrix} \delta A & \delta A & \delta \psi & \delta \psi \end{bmatrix},
\end{equation}
and completing the square to marginalize away $\delta A_H$, $\delta A_L$, $\delta \psi_H$, and $\delta \psi_L$ yields
\begin{gather}
    \ln p \sim - \frac{1}{2}\begin{bmatrix} \delta A - z_A & \delta \psi - z_\psi \end{bmatrix} B \left( 2S - SCS \right) B^\mathrm{T} \begin{bmatrix} \delta A - z_A \\ \delta \psi - z_\psi \end{bmatrix} \\ \left( 2S - SCS\right)B^\mathrm{T} \begin{bmatrix} z_A \\ z_\psi \end{bmatrix} = S\begin{bmatrix} \mu_{A_H} \\ \mu_{A_L} \\ \mu_{\psi_H} \\ \mu_{\psi_L} \end{bmatrix}
\end{gather}
from which we can immediately extract the means and covariances for shared parameters \textit{a posteriori}.
Similarly, we could marginalize away the shared parameters and obtain posteriors for the independent components.
Assuming the observed posteriors are approximately Gaussian for all events, similar analytic marginalization could be used to extract posteriors on shared parameters from a series of events, each sampled independently.

\end{document}